# Policy choices can help keep 4G and 5G universal broadband affordable

Edward J. Oughton[1,3*], Niccolò Comini[2], Vivien Foster[2], Jim W. Hall[3]

[1]George Mason University, Fairfax, VA
[2]World Bank, Washington DC
[3]University of Oxford
*Corresponding author: eoughton@gmu.edu, 4400 University Drive, Fairfax, VA 22030

Abstract

   The United Nations Broadband Commission has committed the international community to accelerate universal broadband. However, the cost of meeting this objective, and the feasibility of doing so on a commercially viable basis, are not well understood. Using scenario analysis, this paper compares the global cost-effectiveness of different infrastructure strategies for the developing world to achieve universal 4G or 5G mobile broadband. Utilizing remote sensing and demand forecasting, least-cost network designs are developed for eight representative low and middle-income countries (Malawi, Uganda, Kenya, Senegal, Pakistan, Albania, Peru and Mexico), the results from which form the basis for aggregation to the global level. The cost of meeting a minimum 10 Mbps per user is estimated at USD 1.7 trillion using 5G Non-Standalone, approximately 0.6% of annual GDP for the developing world over the next decade. However, by creating a favorable regulatory environment, governments can bring down these costs by as much as three quarters – to USD 0.5 trillion (approximately 0.2% of annual GDP) – and avoid the need for public subsidy. Providing governments make judicious choices, adopting fiscal and regulatory regimes conducive to lowering costs, universal broadband may be within reach of most developing countries over the next decade.

1. <u>Introduction</u>

Internet access is increasingly seen as a cornerstone for sustainable development (Bali Swain and Ranganathan, 2021; Hackl, 2018). However, many Low and Middle Income Countries (LMICs) have yet to achieve near-comprehensive population coverage of 3G cellular infrastructure, leading to a growing divide between developed and developing countries (Montenegro and Araral, 2020). While some 4G deployments have taken place (offering peak speeds exceeding 100 Mbps), it can be challenging to deliver significant infrastructure upgrades with substantially lower Average Revenue Per User (ARPU) compared to markets in High Income countries (Hilbert, 2010). Meanwhile, 5G technology is now being rolled-out, bringing substantial performance improvements including speeds of 1 Gbps as well as a new range of uses (Cave, 2018; Oughton et al., 2020). This poses a substantial dilemma – should countries focus on completing (or even beginning) 4G roll-out or 'leapfrog' straight to 5G?

In 2019, the United Nations Broadband Commission called on the international community to work towards an interim milestone of 75 percent global coverage of broadband by 2025. Many countries are now examining how to best encourage the roll-out of universal broadband to support the delivery of the Sustainable Development Goals, but confront the fact that technological and policy decisions involve difficult trade-offs (Forge and Vu, 2020). In making these decisions, government policy makers must grapple with three important choices that determine the speed and cost of meeting access objectives.

First, governments must consider the range of mobile technologies Mobile Network Operators (MNOs) could utilize to meet universal coverage objectives. Although policies are generally technology neutral, governments must consider a range of possible options, such as 4G or 5G, to choose feasible coverage targets reflecting the level of capacity to be provided, the speed of roll-out, and the cost of the network deployment. Second, governments must decide on the desired balance between competition and consolidation in broadband infrastructure development. While competition is generally desirable to create a dynamic and efficient market, in more remote geographic areas demand may not be high enough to support more than one infrastructure provider, and infrastructure sharing facilitated by supportive government regulations may allow network expansion at significantly lower cost. Finally, governments' design of the fiscal regime for broadband infrastructure will affect the viability of roll-out for new generation technologies. Particularly important is the government's policy for pricing access to radio spectrum, which may constitute a significant part of the cost of providing broadband infrastructure. This gives rise to trade-offs between government revenues and population coverage and can lead to tensions between maximizing fiscal revenues from the sector and minimizing the need for public subsidy to reach universal access goals.

The aim of this paper is to estimate the cost of reaching universal access to broadband infrastructure in the developing world, while quantifying how these costs can as far as possible be contained through judicious decisions on technology choice, infrastructure sharing, spectrum pricing and taxation (hence, focusing on supply-side decisions). As the network investments required could indeed be substantial, it is important to undertake independent and transparent assessment of different universal broadband strategies which can improve the viability of high-speed broadband access, potentially using either 4G or 5G. This has already been identified as an important area of research (Ioannou et al., 2020) with the results providing necessary insight for setting national policies for enhancing digital connectivity. The aim is not to necessarily describe the actual dynamics of the roll-out where many cellular technologies may co-exist at the same time, but to examine the general trade-offs in supply-side decisions.

The paper is organized as follows. Section 2 provides an overview of the relevant literature and situates the contribution of this research. Section 3 describes the geospatial simulation methodology used to quantify the costs of alternative broadband access strategies. Section 4 presents and discusses the main findings, and Section 5 concludes.

2. <u>The benefits and challenges of broadband connectivity</u>

A growing number of studies identify the economic benefits of internet access. For example firms which embrace internet access are able to be more productive (Bertschek and Niebel, 2016; Hjort and Poulsen, 2019), more innovative (Paunov and Rollo, 2016) and better at expanding into local and global markets (Gnangnon and Iyer, 2018; Muto and Yamano, 2009). Equally, there are many consumer benefits, which make policies stimulating the adoption of digital technologies attractive for governments. Mobile phone use is positively associated with increased food access (Wantchekon and Riaz, 2019), financial inclusion (Batista and Vicente, 2020; Hasbi and Dubus, 2020), public health care access (Haenssgen et al., 2021; Haenssgen and Ariana, 2017), and poverty reduction (Medeiros et al., 2021; Tadesse and Bahiigwa, 2015). Citizens and governments also gain benefits as cell phones are powerful tools to help reduce corruption (Elbahnasawy, 2014; Kanyam et al., 2017) and improve governance (Asongu and Nwachukwu, 2016).

At the macroeconomic level, better broadband infrastructure is associated positively with a range of macroeconomic indicators, particularly GDP (Briglauer and Gugler, 2019; Czernich et al., 2011; Koutroumpis, 2009). Countries with higher 3G penetration tend to enjoy greater GDP per capita growth than those with comparable total mobile penetration (Williams et al., 2013). Recent estimation suggests that a 10% increase in mobile penetration is associated with an average increase in GDP per capita between 0.59 to 0.76% depending on the model specification (Bahia and Castells, 2019). In the United States, every 10-percentage point gain in penetration annually (from 3G to 4G) has been estimated to generate more than 231,000 jobs (Shapiro and Hassett, 2012). However, broadband roll-out can exhibit diminishing returns to scale, and connection speed is a moderator of this effect, whereby beyond a threshold further quality increases are deemed unproductive (Koutroumpis, 2019). Hence, while digital infrastructure contributes to economic growth, its contribution is far from straightforward and does not necessarily outweigh the high costs of infrastructure roll-out (Tranos, 2012).

Most digital infrastructure around the world is deployed by private network operators via market methods (Cave, 2017; Cave et al., 2019; Gruber, 2005; Jeon et al., 2020; Moshi and Mwakatumbula, 2017; Oughton et al., 2015; Wallsten, 2001; Yoo, 2017). A common challenge in the deployment of digital infrastructure across national territories results from the cost of supply exceeding the price users are willing (or capable) to pay, notably in remote areas (Gerli et al., 2018; Oughton et al., 2018a; Rosston and Wallsten, 2020). Providing access in areas where it may not be commercially viable can be assisted through government policies aimed at reducing the cost of digital infrastructure, and/or provision of government subsidy to make-up the viability gap. A Universal Service Obligation (additionally referred to as a 'coverage obligation') is also a frequent way in which governments mandate private actors to provide service in unviable locations. Such regulatory instruments implicitly mandate the private actor to finance deployment of infrastructure in unviable locations via cross-user subsidization (Curien, 1991), with capital being reallocated from highly profitable areas.

One way that governments can improve the viability of investments to reduce the digital divide is by allowing infrastructure and resource sharing (Afraz et al., 2019; Hou et al., 2019; Maksymyuk et al., 2019; Meddour et al., 2011; Rebato et al., 2016; Samdanis et al., 2016; Sanguanpuak et al., 2019; Van der Wee et al., 2014). The standard presumption that each MNO in the market builds their own dedicated network is increasingly looking unviable, given the cost of delivery in rural areas and the fact revenue is now generally declining in many telecom markets. These considerations become even more relevant with the advent of 5G technology, which is expected to require a large increase in cellular sites to provide greater capacity ('network densification'). Hence, the idea of greater infrastructure sharing for different parts of the network is looking more realistic (Meddour et al., 2011) as MNOs become more open to the idea of market co-operation with competitors (Oughton and Frias, 2018, 2016; Yrjola, 2020), consolidating infrastructure duplication, while producing savings on capital and operational costs (Oladejo and Falowo, 2020).

Spectrum is an essential factor of production for delivering communication services. Efficiently allocating this scarce resource across competing interests is challenging (Gruber, 2005), yet government choices on how it is allocated can influence coverage outcomes. The overall allocation of spectrum must account for the various services which can utilize this resource to maximize aggregate value to society (after deducting non-spectrum supply costs) (Cave and Pratt, 2016). A key challenge of spectrum management is responding to new technologies (e.g. 5G), and the potential institutional changes that may be required (e.g. spectrum sharing and secondary markets) to achieve desirable social outcomes (Gomez et al., 2019; Lehr, 2020; Vuojala et al., 2019; Weiss et al., 2019). Spectrum auctions are a common tool for pursuing efficiency objectives in the cellular market, and the details of their design can be influenced by multiple policy objectives – including revenue mobilization, market efficiency or geographic equity (Goetzendorff et al., 2018). The design of spectrum auctions is often driven primarily by the goal of maximizing fiscal revenues, given the large financial flows that may be involved. However, analysis finds higher spectrum prices are correlated with negative consumer outcomes, including lower coverage levels and slower data speed (Bahia and Castells, 2019). Alternatively, equity issues can be addressed by imposing geographic or population coverage obligations as part of the auction design, accepting that this will lead to lower fiscal revenues (Cave and Nicholls, 2017), but potentially higher net social welfare (Björkegren, 2019). Auction design can also be used to address excessive market power by setting aside spectrum for new entrants, or by limiting the aggregate amount of spectrum a single MNO can hold (Peha, 2017).

Where neither infrastructure sharing or spectrum auction design are enough to close the viability gap for universal broadband service, governments may decide to provide subsidies to encourage availability in unviable locations (Bourreau et al., 2020). Governments have subsidized the roll-out of telecommunication infrastructure for decades, therefore many assessments have addressed this issue, whether for fixed telephone lines or high-speed broadband (Kenny and Kenny, 2011; Prieger, 2013; Rajabiun and Middleton, 2014).

Governments also need to consider the trade-off between taxing the mobile sector while at the same time directly providing subsidies in unviable locations to achieve universal access. Due to network externalities of connecting to mobile phone networks, the social welfare cost of taxing the mobile sector may be as much as three times the fiscal revenue raised (Björkegren, 2019).

3. Method

A new high-resolution simulation model is developed to assess the viability of different strategic policy choices relating to 4G and 5G network roll-out in emerging economies. The simulation method is based on scenario analysis, which enables 'what-if' questions to be tested (Chen et al., 2020; Jefferson, 2020; Kayser and Shala, 2020; Kishita et al., 2020; Meadows and O'Brien, 2020). Scenario analysis is a common approach to explore technology futures (where historical experience is not a dependable guide to the future) and provides useful insight into the comparative performance of different decision options, especially for policy (Crawford, 2019; Frith and Tapinos, 2020; Gordon, 2020; Hutajulu et al., 2020; Metz and Hartley, 2020; Wright et al., 2020).

The method consists of three steps. First, countries with similar demand-side and supply-side broadband characteristics are clustered. Second, a 4G and 5G assessment model capable of simulating roll-out strategies is developed to test the implications of different decisions in each cluster of countries. Finally, the cost per capita for the specific countries modeled is generalized to all other nations in each cluster, based on their population.

3.1 Country clustering

Detailed modeling of all 135 emerging economies would be extremely challenging. Therefore, the motivation for clustering countries is to allow a subset of representative countries to be assessed for universal broadband deployment. Grouping by cluster then enables the cost per population from the assessed countries to be generalized to other countries within that cluster, reducing estimation uncertainty.

Countries are clustered based on a combination of factors which affect digital infrastructure viability. Using a theoretical understanding of broadband adoption and infrastructure deployment from the literature, three main factors are selected (Gallardo et al., 2020; Koutroumpis, 2009; Reddick et al., 2020; Rhinesmith et al., 2019). The demand-side is captured by Gross Domestic Product (GDP) per capita, as an indicator of the ability to pay for broadband services such as 4G or 5G, and population density, as an indication of the maximum possible demand available in any geographic area. The supply-side is represented by existing 4G network coverage, which reflects the current level of infrastructure provision and therefore future incremental cost.

High income countries are excluded from the clustering process, as are outliers with very extreme values (such as The Maldives) to avoid outliers skewing the clustering results. This allows the analysis to focus on developing countries where the challenges of broadband deployment are driven by the need for wide-area coverage in areas of low population density and/or low take-up. Using K-means clustering, a popular classical unsupervised machine learning method, six clusters are exogenously specified. This quantity significantly reduces the group-level variation via the Within-Group Sum of Squares (Hothorn and Everitt, 2014), while making the results easy to obtain and report. The default R *stats* algorithm defined by Hartigan and Wong (Hartigan and Wong, 1979) is applied. Data on GDP per capita and population density (World Bank, 2020), as well as 4G coverage (GSMA, 2020), enable Figure 1 to report the summary cluster statistics, along with a global map indicating the country clusters. A detailed description of the clustering process is provided in Section 1 of the Supplementary Materials.

Figure 1 Clustering summary statistics and global mapping

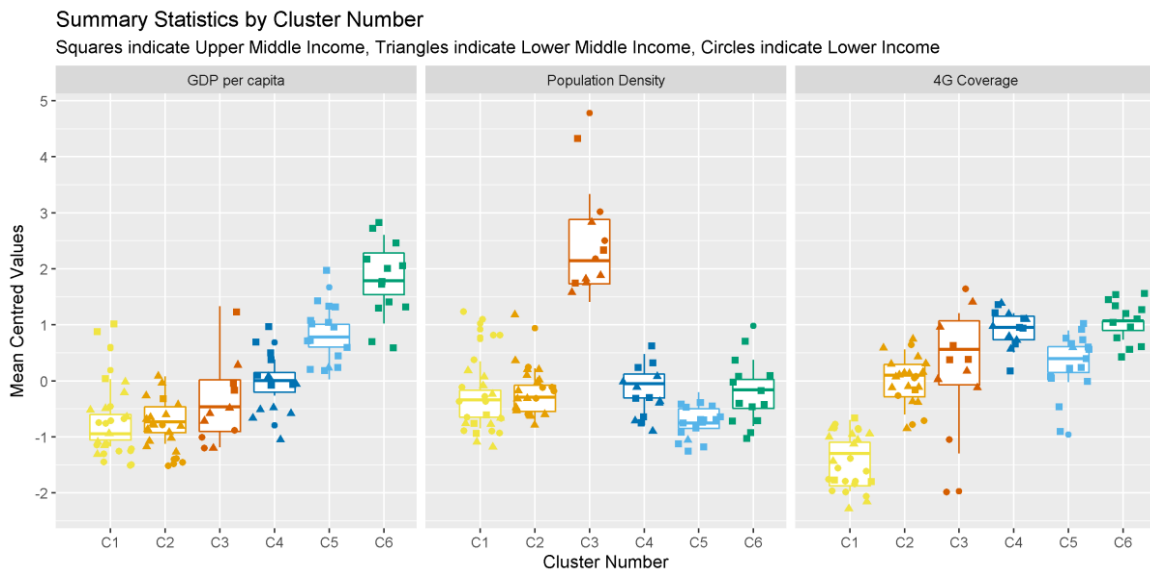

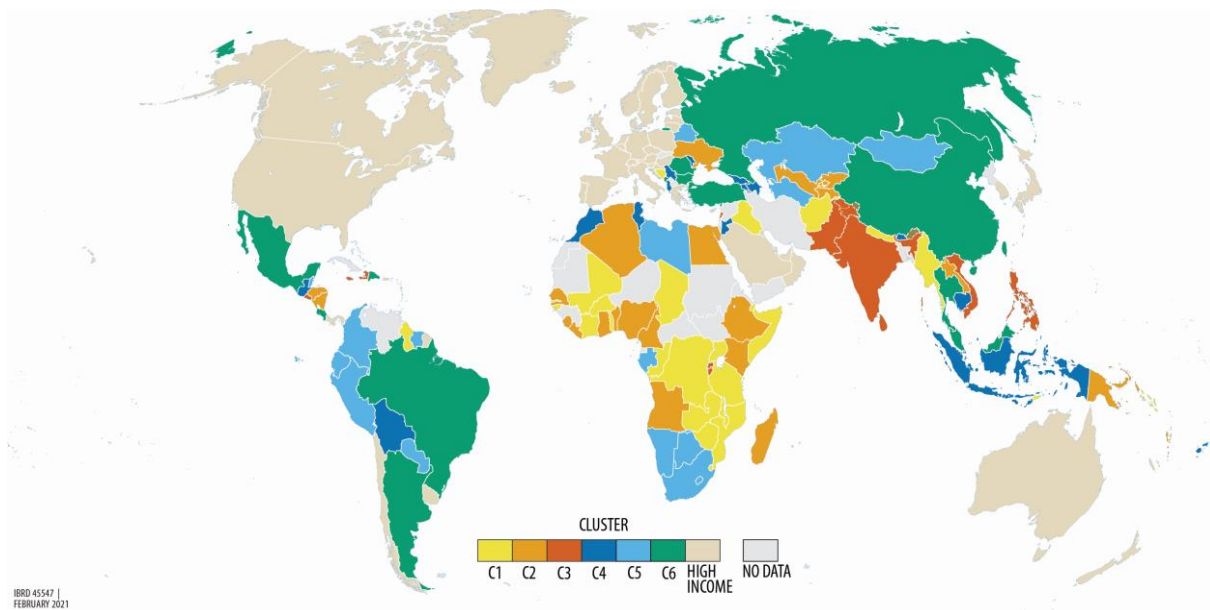

Cluster 1 (C1) has mainly low-income countries with below average population density and 4G coverage. Cluster 2 (C2) also has mainly low-income countries with below average population density, but above average 4G coverage. Cluster 3 (C3) includes very densely populated countries, with low GDP per capita and 4G coverage only marginally above average. Cluster 4 (C4) includes middle-income countries with mean population density, but above average 4G coverage. Cluster 5 (C5) also comprises middle-income countries with above average 4G coverage, but below average population density. Finally, Cluster 6 (C6) includes mainly upper middle-income countries, with above average 4G coverage, and mean population density. Representative countries are then selected from each cluster ensuring a reasonable geographic spread and the practicalities of data availability. These include Uganda and Malawi from Cluster 1, Kenya and Senegal from Cluster 2, Pakistan from Cluster 3, Albania from Cluster 4, Peru from Cluster 5, and Mexico from Cluster 6. The key characteristics of these countries are shown in Table 1, using metrics sourced from the World Bank (World Bank, 2020), GSMA (GSMA, 2020) and where available crowdsourced mobile broadband results (Speedtest, 2020). Generally, these countries differ from left to right in terms of development, so Malawi has the lowest income per capita, whereas Mexico has the highest. This effect subsequently has an impact on 4G coverage, which generally increases from left to right, as the cluster numbers increase.

Table 1 Selected countries for analysis

| Cluster | C1 | C1 | C2 | C2 | C3 | C4 | C5 | C6 |
|---|---|---|---|---|---|---|---|---|
| Country | Malawi | Uganda | Senegal | Kenya | Pakistan | Albania | Peru | Mexico |
| Region | SSA | SSA | SSA | SSA | SA&SE | E&CA | LAC | LAC |
| Income group | Low | Low | Lower-middle | Lower-middle | Lower-middle | Upper-middle | Upper-middle | Upper-middle |
| 4G Coverage | 16% | 31% | 50% | 61% | 67% | 96% | 84% | 85% |
| Population Density (per km2) | 192 | 213 | 82 | 90 | 275 | 105 | 25 | 65 |
| Population (m) | 18.1 | 42.7 | 15.9 | 51.4 | 212.2 | 2.9 | 32.0 | 126.2 |
| GDP Per Capita | USD 389 | USD 643 | USD 1,522 | USD 1,711 | USD 1,473 | USD 5,254 | USD 6,947 | USD 9,698 |
| Rural Population | 83% | 76% | 53% | 73% | 63% | 40% | 22% | 20% |
| Area (km2) | 94,280 | 200,520 | 192,530 | 569,140 | 770,880 | 27,400 | 1,280,000 | 1,943,950 |
| Mean Mobile Broadband Speedtest (Mbps) | - | 11 | 22 | 22 | 17 | 51 | - | 30 |

3.2 Python Telecommunication Assessment Library (*pytal*)

While there have been analyses of these selected countries focusing on various parts of the digital ecosystem (Ahmad et al., 2019; Avilés, 2020; Cave and Mariscal, 2020; Hameed et al., 2018; Ignacio et al., 2020; Mir and Dangerfield, 2013; Ovando, 2020), few examples develop a scientifically reproducible open-source codebase which can be used to test real world decisions. Building on national 5G assessments carried out in OECD markets, such as the UK (Oughton et al., 2018b; Oughton and Frias, 2016, 2018; Oughton and Russell, 2020) and the Netherlands (Oughton et al., 2019a), the Python Telecommunications Assessment Library (pytal) is developed to provide a globally-scalable simulation model which can be applied to undertake cross-country comparative 4G and 5G analytics. Figure 2 visualizes the assessment method.

Figure 2 National 4G and 5G assessment method in *pytal*

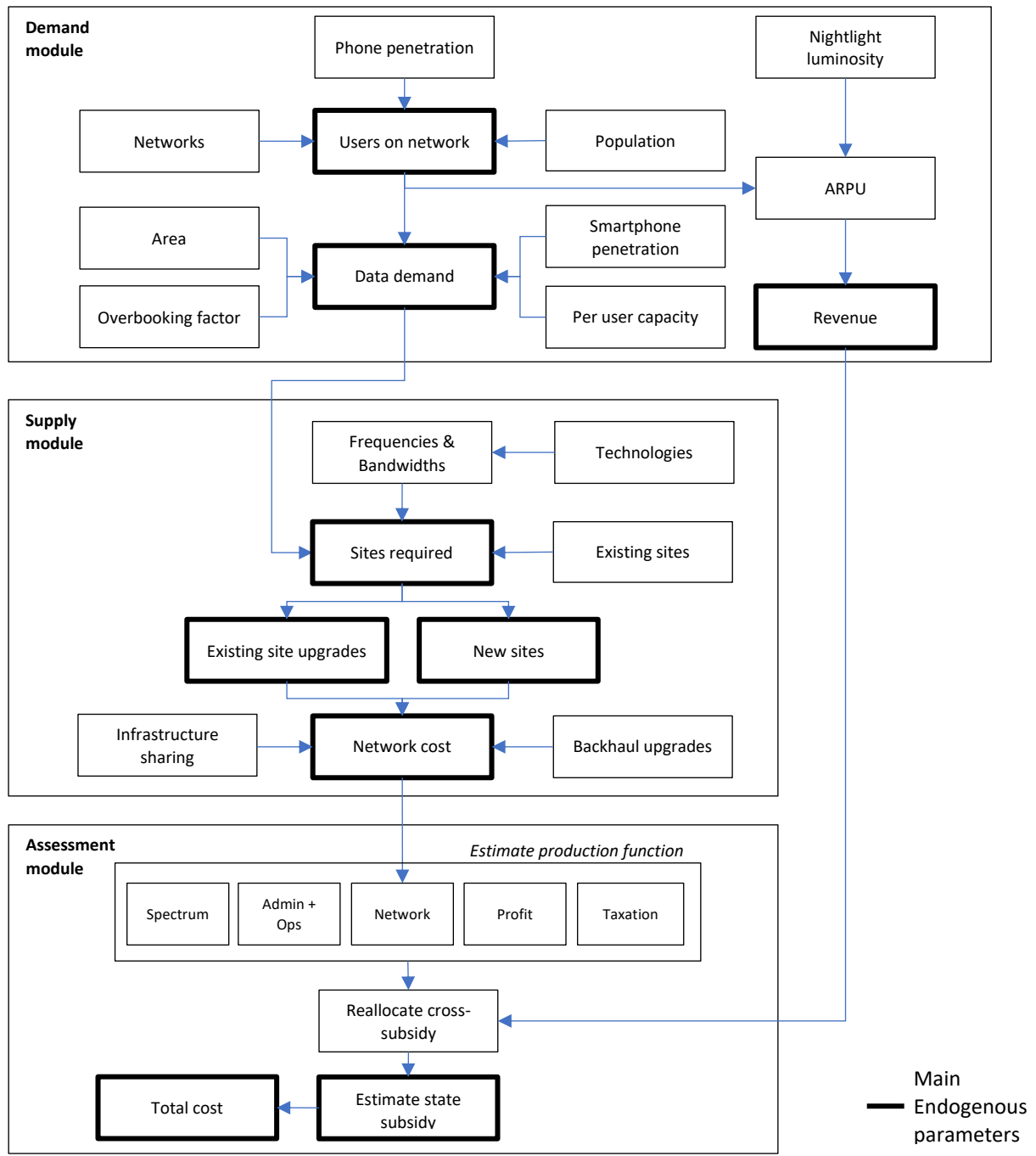

### 3.2.1 Demand module

The demand module follows a standard infrastructure forecasting method (Thoung et al., 2016). The density of the population with cell phones ($Users\_on\_network_{it}$) in the $i$th area per km² for time period $t$ is calculated using the local population density ($Population_i$), the percentage of unique cell phone users ($Penetration_{it}$) and the number of cellular networks ($Networks_i$), as illustrated in equation (1):

$$Users\_on\_network_{it} = \frac{Population_i \cdot (Penetration_{it} / 100)}{Networks_i} \quad (1)$$

Spatial population estimates are obtained using a global 1 km² gridded dataset (Tatem, 2017; WorldPop, 2019). Penetration data for unique cellular subscribers is taken from the GSMA for 2010-2020 (GSMA, 2020) as illustrated in Figure 3 (A). The number of networks in the user calculation is exogenously set depending on the present number of national networks in each market, yielding a representative market share for a hypothetical operator.

The data demand ($Demand_{it}$) in the $i$th area (km²) needing to be met from the number of users on the network ($Users\_on\_network_i$) in time period $t$ can be obtained by taking the smartphone penetration rate ($SPPenetration_{it}$), the desired per user capacity ($Scenario$) and an overbooking factor ($OBF$) as not all users will access the network at the same time, as illustrated in equation (2):

$$Demand_{it} = \frac{Users_i \cdot SPPenetration_{it} \cdot Scenario}{OBF} \quad (2)$$

Historical data are used up to 2020, with future years forecast at a compound rate coherent with the historical adoption trajectory (Malawi: 3.5%, Uganda: 2%: Senegal: 1.5%, Kenya: 1.5%, Pakistan: 2.5%, Albania: 0.3%, Peru: 0.5%, Mexico: 1.2%). The number of networks in the user calculation is exogenously set depending on the present number of national networks in each market, yielding a representative market share for a hypothetical operator.

Figure 3 Demand forecasts by country

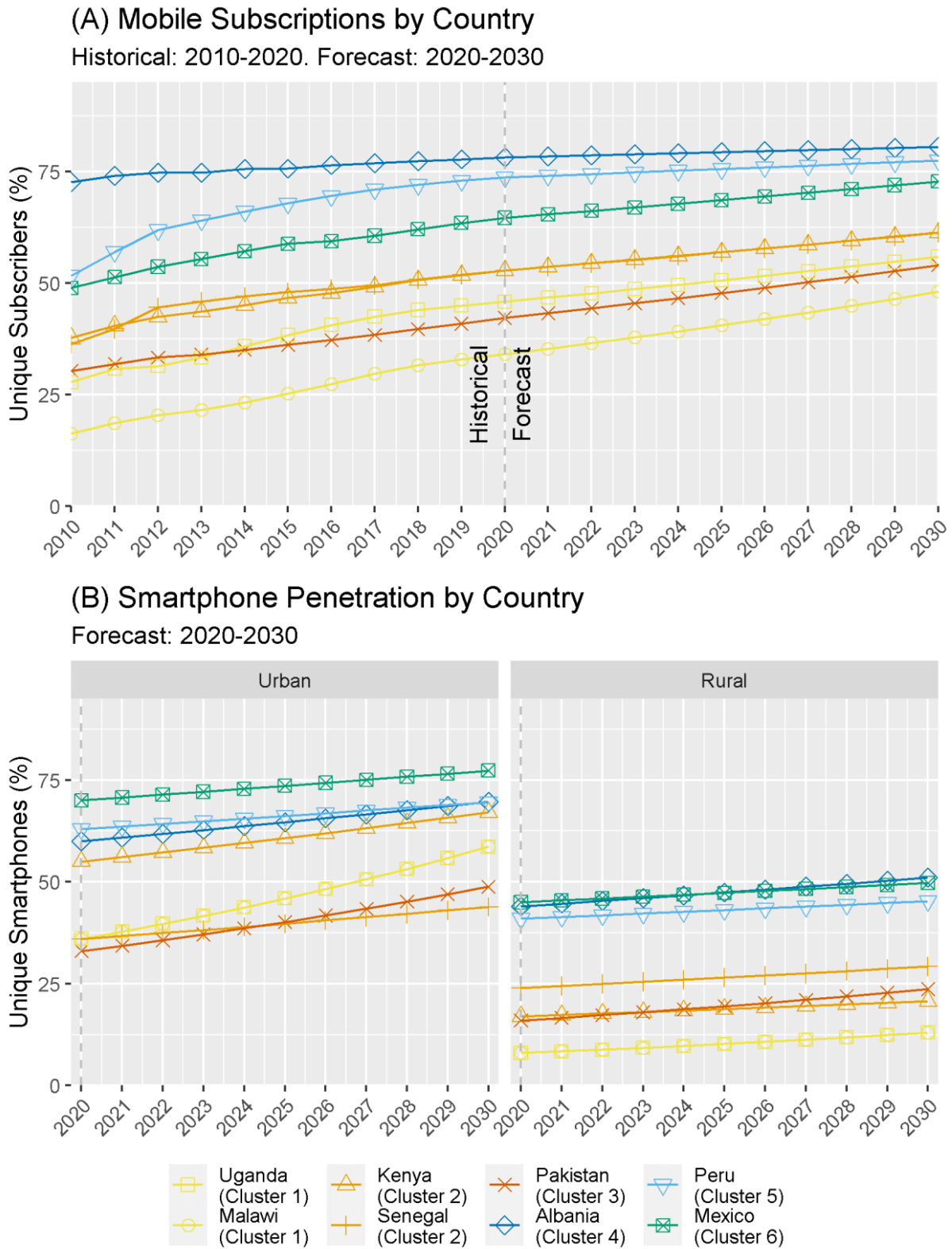

Figure 3 (B) contains the smartphone adoption rates for urban and rural users, based on survey data (Research ICT Africa, 2018) forecast forward over the study period annually (equating to national smartphone growth as follows: Malawi: 5%, Uganda: 5%: Senegal: 2%, Kenya: 2%, Pakistan: 4%, Albania: 1.5%, Peru: 1%, Mexico: 1%). With cellular networks being designed for either capacity, coverage, or a mixture of the two, the scenarios tested here take this into account. To ensure viability, high capacity is targeted at users in urban and suburban areas as cities form only a small proportion of total land mass, with lower capacity targeted at rural areas. In Scenario 1 (S1), users are targeted with 25 Mbps in urban, 10 Mbps in suburban and 2 Mbps in rural. In Scenario 2 (S2), users are targeted with 200 Mbps in urban, 50 Mbps in suburban and 5 Mbps in rural. In Scenario 3 (S3), users are targeted with 400 Mbps in urban, 100 Mbps in suburban and 10 Mbps in rural. It should be noted that S3 is the only scenario that fully meets the goal of the UN Broadband Commission, which suggests a minimum capacity of 10 Mbps for the entire population. For QoS an overbooking factor of 20 is used ($OBF$ = 20), meaning from twenty users only one user is trying to access the available network resources at any one time.

The expected revenue ($Revenue_{it}$) in the $i$th area per km² can then be estimated given the Average Revenue Per User ($ARPU_i$), the density of cell phone users ($Users_{it}$), over the time horizon of the analysis in years ($TimeHorizon$), as illustrated in equation (3):

$$Revenue_{it} = \sum ARPU_i \cdot Users_{it} \cdot 12 \qquad (3)$$

The revenue is discounted over the ten-year assessment period using a discount rate of 5%. The ARPU is estimated based on extracted nighttime luminosity values which are averaged over each region. Luminosity data are available in Digital Number (DN) units. High luminosity is treated as being >20 DN km² on average, medium luminosity as >15 DN km² on average, and low luminosity as being <15 DN km² on average. The monthly ARPU amounts allocated to each of these bands by country are listed in Table 2, based off the average 2020 ARPU (GSMA, 2020).

Table 2 Monthly ARPU by country

| Type | Monthly ARPU ($) | | | | | | | |
|---|---|---|---|---|---|---|---|---|
| | Malawi | Uganda | Senegal | Kenya | Pakistan | Albania | Peru | Mexico |
| High | 3.5 | 3.5 | 8 | 8 | 4 | 8 | 11 | 11 |
| Medium | 2.5 | 2.5 | 5 | 5 | 2 | 7 | 10 | 10 |
| Low | 1 | 1 | 2 | 2 | 1 | 4 | 5 | 5 |

### 3.2.2 Supply module

The supply module is based on a geospatial system model which can estimate the number of additionally required sites in order to meet the demand present in an area. The area capacity is estimated by obtaining the mean Network Spectral Efficiency ($\bar{\eta}_{area}^{f}$) (bps/Hz/km2) given the average number of cells per site ($\bar{\eta}_{cells}$) and density of co-channel sites ($\rho_{sites}$) using the same spectrum frequency ($f$), as detailed in eq. (4).

$$\bar{\eta}_{area}^{f} = \bar{\eta}_{cells}^{f} \cdot \rho_{sites} \qquad (4)$$

The total capacity for an area ($Capacity_{area}$) can then be estimated for all frequencies by multiplying the Network Spectral Efficiency ($\bar{\eta}_{area}^{f}$) by the available spectrum bandwidth ($BW^{f}$), as in eq. (5).

$$Capacity_{area} = \sum_{f} \bar{\eta}_{area}^{f} \; BW^{f} \qquad (5)$$

The technical engineering method which underpins the supply module can be found in Section 2 in the Supplementary Materials. This includes comprehensive detail on the estimation of the baseline infrastructure, least-cost network designs, upgrade strategies, costs and spectrum prices. Figure 4 provides illustrative examples of the least-cost fiber network designs, where black connections represent existing fiber, red connections represent new fiber and yellow nodes represent are new fiber access points.

Figure 4 Least-cost fiber network designs for existing and planned connections

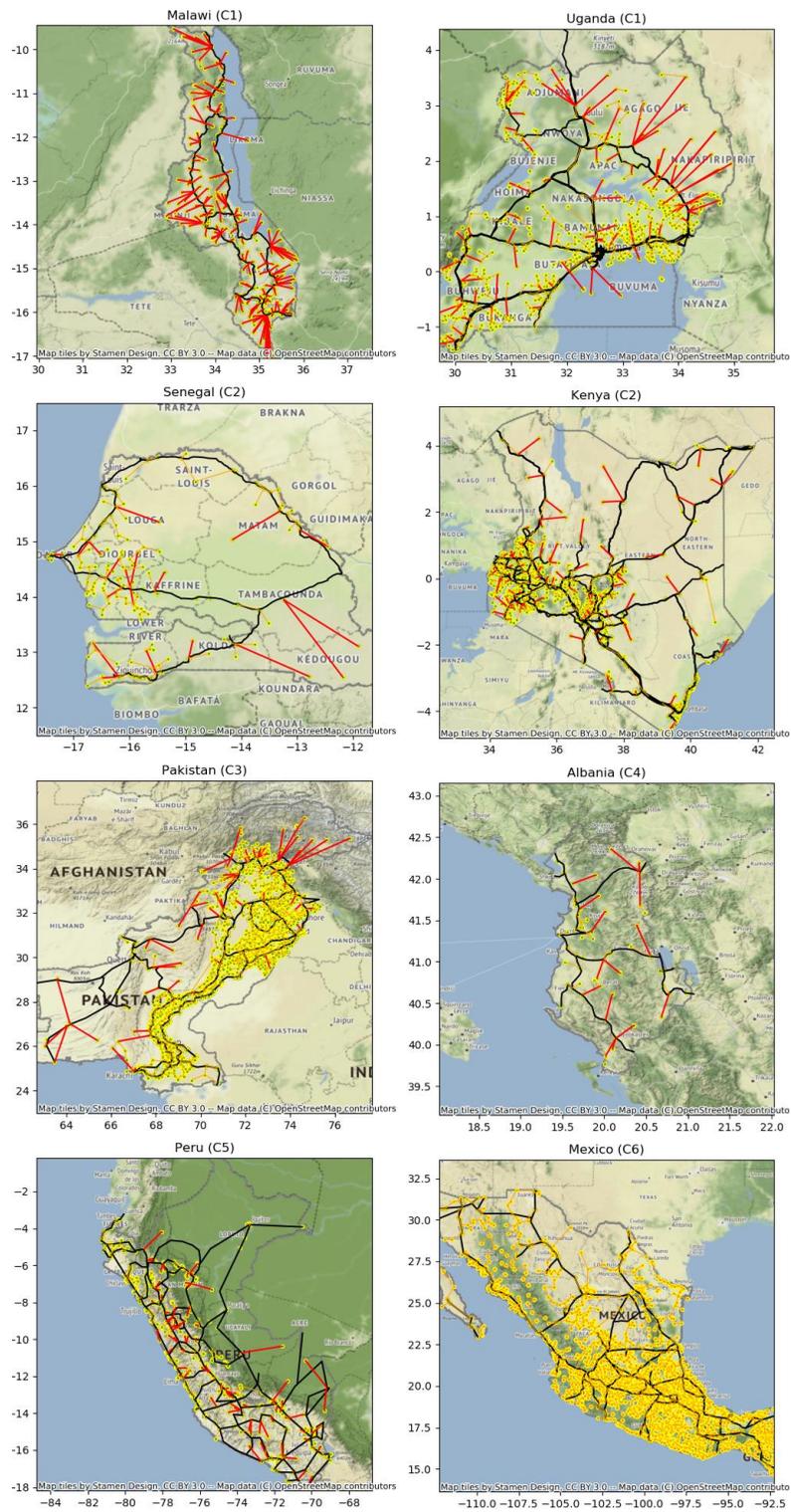

Black: Existing network, Red: required network build-out, Yellow: Regional fiber edges and nodes.

### 3.2.3 Cost estimation

The cost module first calculates the total discounted cost of capital and operating expenditures over the 10-year horizon, to obtain the Net Present Value in the first year of the assessment period (2020). Based on a highly detailed cost assessment reported in Section 2.6 of the Supplementary Materials, a typical three-sector macro cell requiring three remote radio units has a greenfield cost of $31.5k for equipment, $20k for the site build, $5k for installation, which is broadly similar with the stated costs in the literature. As administration is added later, these costs are lower than other studies that may use site costs between $100-200k each. The cost of electricity for each macro cell annually varies but is treated as $5k per year for 4G and $10k for 5G, which is much higher than European-centric estimates (METIS, 2017) but accounts for higher local energy prices, whilst also reflecting the fact that 5G consumes far more power than 4G. Other operating costs for maintenance and servicing are treated as being 10% of the investment capex of all active electronic components (METIS, 2017), similar to other estimates of 5-10% (Markendahl and Mäkitalo, 2010). In terms of other annual operating costs, rental would be between $1-9.6k for the site depending on the settlement type (urban, suburban or rural) and maintenance costs would be $3.15k, producing an annual cost of ~$8.8k excluding the backhaul. The backhaul, regional and core fiber network is built and operated by the MNO and is dependent on the geographic area it needs to cover, given the least-cost network structure. Hence, rural costs are generally higher as links need to traverse longer distances.

An administration cost is also used, which can be as high as 30% in OECD nations (Rendon Schneir et al., 2019), but this value is adapted for countries with significantly lower labor prices, hence 20% of the network cost is applied annually. A Weighted Average Cost of Capital (WACC) is applied at 15% (WACC Expert, 2020), which is substantially higher than OECD nations, but this value is set by money markets and reflects the relative risk of lending (which can be higher in some countries due to political stability, poorer legal transparency, corruption etc.).

3.2.4 Spectrum costs

Spectrum costs consist of an upfront reserve price, a competitive premium and any associated annual fees over the license duration. Identifying spectrum costs for all countries globally can be challenging as not all information is made public. A sensible approach for the baseline is to use past spectrum prices to guide approximate values, after controlling for bandwidth, population and other major factors that make international comparison challenging. Based on statistical distribution for historical auction prices paid for different spectrum categories based on the six cluster groupings identified, Section 2.7 of the Supplementary Materials presents the assumed dollar cost per MHz per head of the population, adjusting for frequency bandwidth and market potential. The results for each cluster are shown in Table 3, indicating a factor of 10 in the variation of spectrum costs across countries, illustrating the sensitivity of these costs to market conditions and local government policies.

Table 3 Baseline spectrum prices by country

| Type | Spectrum cost ($/MHz/Pop) | | | | | | | |
|---|---|---|---|---|---|---|---|---|
| | Malawi | Uganda | Senegal | Kenya | Pakistan | Albania | Peru | Mexico |
| Coverage | 0.02 | 0.02 | 0.15 | 0.15 | 0.05 | 0.40 | 0.20 | 0.20 |
| Capacity | 0.01 | 0.01 | 0.08 | 0.08 | 0.03 | 0.10 | 0.10 | 0.10 |

3.3 Assessment of government policy choices

The simulation model is used to estimate the baseline cost of reaching universal access to broadband in the selected countries, and further to assess the implication of alternative government policy choices with regard to the selected generation of mobile technology, the extent of infrastructure sharing, and the pricing of spectrum.

When it comes to mobile technology choice, three main cellular technologies are included in the upgrade strategies tested, each of which deliver different use cases. Upgrading to 4G simply provides

Mobile Broadband. On the other hand, upgrading to 5G non-standalone (NSA) delivers both Enhanced Mobile Broadband and Massive Machine Type Communications, while going all the way to 5G standalone (SA) additionally provides Ultra-Reliable and Low-Latency Communications. Two backhaul technologies are tested for both 4G and 5G technologies, including fixed fiber optic and wireless.

In terms of the regulatory environment for infrastructure sharing, several options are considered. Under the baseline scenario, each MNO builds its own dedicated network. In passive site sharing, MNOs share physical facilities, such as the site compound and tower, but not any electrical components. With a Multi Operator Radio Access Network (MORAN), known as active sharing, MNOs share all passive and active components, including the radio heads and backhaul. The most complete form of infrastructure sharing is to mandate a single Shared Rural Network (SRN), which allows MNOs both to share all passive and active equipment via a MORAN, as well as a shared core and regional fiber network only in rural areas.

As far as spectrum pricing is concerned, changes from the baseline spectrum prices reported above are explored via a -75% and +100% change in cost to provide low and high scenarios respectively. Corporate taxation is also incorporated into the model at an illustrative rate of 30%. Although simulations were also conducted for low (10%) and high (40%) variations of taxes, results are not reported due to limited sensitivity.

For each scenario, the model reports the percentage of the population that can be covered on a commercially viable basis, after accounting for excess profits (>20%) being reallocated via user cross-subsidization to unviable locations. It also reports the total social cost of meeting 100% coverage, including the public subsidy that may be required to close any remaining viability gap.

The total social cost can be obtained by the summation of the total cost to the network operator ($Operator\_Cost_i$) and government ($Government\_Cost_i$), as detailed in eq. (6).

$$Societal\_Cost_i = Government\_Cost_i + Private\_Cost_i \tag{6}$$

The government cost ($Government\_Cost_i$) can be calculated based on any subsidies required to roll out infrastructure in unviable locations ($Subsidy_i$), minus any additional fiscal revenue obtained as a result of the network roll-out, via either spectrum license fees ($Spectrum_i$) and taxation ($Tax_i$) as detailed in eq. (7).

$$Government\_Cost_i = Subsidy_i - (Spectrum_i - Tax_i) \tag{7}$$

The operator cost ($Private\_Cost_i$) for the $i$th area can be obtain by taking the summation of the network cost ($Network_i$), administration cost ($Administration_i$), spectrum cost ($Spectrum_i$), tax ($Tax_i$), plus the profit margin ($Profit_i$), as illustrated below in equation (8):

$$Private\_Cost_i = Network_i + Administration_i + Spectrum_i + Tax_i + Profit_i \tag{8}$$

3.4 Model uncertainty and limitations

The model presented takes a cost-minimization approach but does not extend to include a full behavioral model of economic agents. For example, on the supply-side agents may respond to changes in relative prices depending on different market structure and electricity access scenarios, whereas on the demand-side different education and age group demographics could affect take-up.

Whilst great care is taken to obtain accurate estimates of model inputs, there are several inevitable sources of uncertainty. Uncertainty is introduced by not having explicit geospatial information on (i) cellular sites or regional fiber, (ii) ARPU, phone penetration or smartphone ownership, and (iii) backhaul technology. The methods used to derive these estimated data layers ensure that the aggregate quantity of supply-side infrastructure or demand-side users is accurate (e.g. total towers, total users etc.), but uncertainty is introduced in the regional spatial allocation of assets and users in the assessment.

Additionally, in the absence of other evidence, the modeling assumes that spectrum prices for 5G, which are a major cost driver, will be like those applied for previous generation technologies. For example, higher capacity 5G services depend on larger spectrum bandwidth. However, historical spectrum prices relate to much smaller bandwidths (e.g. 10 MHz bandwidth at 800 MHz for 4G) than the larger spectrum bandwidth needed to support much higher capacity 5G services. In that sense, historic prices may not be a reliable guide to what would emerge in future 5G auctions.

Finally, the simulations assume that MNOs can take a fair profit margin based on the capital employed to build the network (e.g. 20%). This is in addition to the WACC risk premium and reflects the risk involved with infrastructure investment. Higher revenues in areas with high demand are reallocated to unviable locations, essentially forcing some degree of geographic cross-subsidization, and thereby reducing the viability gap relative to the outcome of a pure profit-maximizing strategy.

4. Results and Discussion

This section presents the results of the simulation exercise described above for the study period from 2020-2030. As the policy objective of providing universal access to broadband has already been identified, the analysis focuses on how this can be reached in the most cost-effective way. In making these judgments, the total social cost is used as the primary performance metric, which is to say the industry's production cost net of taxation but inclusive of any government subsidies. The analysis also explores the net fiscal impact of achieving universal access, considering both government revenues from spectrum pricing and corporate taxations, as well as government subsidies to the sector where these are warranted to cover remaining viability gaps. The discussion of the results also considers how these can be used to inform government policy choices on technology, the design of the regulatory framework for infrastructure sharing, and the pricing of spectrum.

### 4.1 Social cost of universal access

The cost of reaching broadband universal service varies dramatically according to the technology selected and the desired level of capacity (see Table 4).

For any level of capacity demanded per user, the differential between the most expensive technology choice compared to the most cost-effective one is several multiples. The magnitude of the cost differential between technologies generally increases with the baseline level of 4G coverage. Countries like Malawi and Uganda (with 4G coverage of 15-30%) face cost differentials of 200-400% between technology choices, whereas countries like Albania and Mexico (with 4G coverage of 85-95%) can face cost differentials as high as 1000% between different technologies, driven by the existing level of 4G investment. For 4G technology, the consequence of choosing fiber backhaul is to almost double the cost of meeting universal access, although this differential drops at higher levels of user capacity demand. In the case of 5G technology, the choice of fiber backhaul at least doubles the cost of reaching universal coverage in most cases.

Table 4 Social cost results for all technologies, scenarios and strategies (NPV 2020-2030)

Technology Results reported by Country

| | | | C1 | | C2 | | C3 | C4 | C5 | C6 |
|---|---|---|---|---|---|---|---|---|---|---|
| Scenario | Strategy | Metric | Malawi | Uganda | Senegal | Kenya | Pakistan | Albania | Peru | Mexico |
| S1 (<25 Mbps) | 4G (W) | Social Cost ($Bn) | 1.2 | 1.4 | 0.96 | 3.3 | 18 | 0.023 | 2.9 | 5.8 |
| S1 (<25 Mbps) | 4G (F) | Social Cost ($Bn) | 1.9 | 3.3 | 1.7 | 5.6 | 32 | 0.031 | 5 | 10 |
| S1 (<25 Mbps) | 5G NSA (W) | Social Cost ($Bn) | 1.5 | 1.4 | 0.87 | 2.9 | 12 | 0.19 | 6.2 | 35 |
| S1 (<25 Mbps) | 5G SA (F) | Social Cost ($Bn) | 4.1 | 8.8 | 2.2 | 9 | 58 | 0.54 | 15 | 71 |
| S2 (<200 Mbps) | 4G (W) | Social Cost ($Bn) | 3 | 3.4 | 1.7 | 6.3 | 81 | 0.046 | 5.7 | 15 |
| S2 (<200 Mbps) | 4G (F) | Social Cost ($Bn) | 3.8 | 7 | 2.7 | 11 | 120 | 0.066 | 8.7 | 21 |
| S2 (<200 Mbps) | 5G NSA (W) | Social Cost ($Bn) | 2.1 | 3 | 1.6 | 4.9 | 53 | 0.25 | 8.6 | 48 |
| S2 (<200 Mbps) | 5G SA (F) | Social Cost ($Bn) | 5 | 14 | 4.5 | 14 | 310 | 0.61 | 34 | 130 |
| S3 (<400 Mbps) | 4G (W) | Social Cost ($Bn) | 5.2 | 7.2 | 2.7 | 13 | 160 | 0.25 | 9.4 | 19 |
| S3 (<400 Mbps) | 4G (F) | Social Cost ($Bn) | 6.2 | 12 | 5.1 | 20 | 210 | 0.34 | 14 | 27 |
| S3 (<400 Mbps) | 5G NSA (W) | Social Cost ($Bn) | 2.8 | 4.1 | 1.9 | 6.1 | 82 | 0.32 | 11 | 54 |
| S3 (<400 Mbps) | 5G SA (F) | Social Cost ($Bn) | 5.9 | 15 | 5.2 | 18 | 560 | 0.71 | 67 | 150 |

[1] Infrastructure Sharing Strategy: Baseline.
[2] Spectrum Pricing Strategy: Baseline.
[3] Taxation Strategy: Baseline.
[4] Results rounded to 2 s.f.

The desired capacity per user evidently also has a significant impact on total costs of meeting universal access, although unit costs decline dramatically. Considering 5G (NSA) technology, the cost of providing the higher capacity represented by S3 (<400 Mbps) is in most cases approximately double the cost associated with providing capacity of S1 (<25 Mbps). However, the capacity provided to users under S3 is up to 16 times higher than S1, indicating the magnitude of scale economies in the provision of capacity. These results indicate the importance of having a clear understanding of likely growth in capacity demanded by users when designing a broadband network.

### 4.2 Impact of technology choice

Governments intent on minimizing the cost of meeting universal access goals for broadband must identify the most relevant technology for any specific country setting. Table 5 presents the lowest social cost technology in each country across all scenarios for capacity per user and encompassing policy choices on infrastructure sharing and spectrum pricing. Several striking findings emerge.

First, for all but two countries (Malawi and Uganda), the same technology is always least cost in each country, irrespective of the demand scenario or the adoption of different policies to reduce costs. This suggests that underlying country characteristics, such as economic geography and legacy infrastructure, are the critical determinants of technology choice. Only in the case of Malawi and Uganda, does the attractiveness of a specific technology choice depend on higher levels of capacity demanded per user.

Table 5 Identifying the cheapest technology for achieving universal service

Least (Social) Cost Technology for Universal Coverage

| Country | Scenario | Baseline | Infrastructure Sharing | | | Spectrum Pricing | | Taxation | | Hybrid |
| --- | --- | --- | --- | --- | --- | --- | --- | --- | --- | --- |
| | | | Passive | Active | SRN | Low P. | High P. | Low T. | High T. | Lowest |
| Malawi | S1 (<25 Mbps) | 4G (W) | 4G (W) | 4G (W) | 4G (W) | 4G (W) | 4G (W) | 4G (W) | 4G (W) | 4G (W) |
| Malawi | S2 (<200 Mbps) | 5G NSA (W) | 5G NSA (W) | 5G NSA (W) | 5G NSA (W) | 5G NSA (W) | 5G NSA (W) | 5G NSA (W) | 5G NSA (W) | 5G NSA (W) |
| Malawi | S3 (<400 Mbps) | 5G NSA (W) | 5G NSA (W) | 5G NSA (W) | 5G NSA (W) | 5G NSA (W) | 5G NSA (W) | 5G NSA (W) | 5G NSA (W) | 5G NSA (W) |
| Uganda | S1 (<25 Mbps) | 4G (W) | 4G (W) | 4G (W) | 4G (W) | 4G (W) | 4G (W) | 4G (W) | 4G (W) | 4G (W) |
| Uganda | S2 (<200 Mbps) | 5G NSA (W) | 5G NSA (W) | 4G (W) | 5G NSA (W) | 5G NSA (W) | 5G NSA (W) | 5G NSA (W) | 5G NSA (W) | 5G NSA (W) |
| Uganda | S3 (<400 Mbps) | 5G NSA (W) | 5G NSA (W) | 5G NSA (W) | 5G NSA (W) | 5G NSA (W) | 5G NSA (W) | 5G NSA (W) | 5G NSA (W) | 5G NSA (W) |
| Senegal | S1 (<25 Mbps) | 5G NSA (W) | 5G NSA (W) | 4G (W) | 5G NSA (W) | 5G NSA (W) | 5G NSA (W) | 5G NSA (W) | 5G NSA (W) | 5G NSA (W) |
| Senegal | S2 (<200 Mbps) | 5G NSA (W) | 5G NSA (W) | 5G NSA (W) | 5G NSA (W) | 5G NSA (W) | 5G NSA (W) | 5G NSA (W) | 5G NSA (W) | 5G NSA (W) |
| Senegal | S3 (<400 Mbps) | 5G NSA (W) | 5G NSA (W) | 5G NSA (W) | 5G NSA (W) | 5G NSA (W) | 5G NSA (W) | 5G NSA (W) | 5G NSA (W) | 5G NSA (W) |
| Kenya | S1 (<25 Mbps) | 5G NSA (W) | 5G NSA (W) | 5G NSA (W) | 5G NSA (W) | 5G NSA (W) | 5G NSA (W) | 5G NSA (W) | 5G NSA (W) | 5G NSA (W) |
| Kenya | S2 (<200 Mbps) | 5G NSA (W) | 5G NSA (W) | 5G NSA (W) | 5G NSA (W) | 5G NSA (W) | 5G NSA (W) | 5G NSA (W) | 5G NSA (W) | 5G NSA (W) |
| Kenya | S3 (<400 Mbps) | 5G NSA (W) | 5G NSA (W) | 5G NSA (W) | 5G NSA (W) | 5G NSA (W) | 5G NSA (W) | 5G NSA (W) | 5G NSA (W) | 5G NSA (W) |
| Pakistan | S1 (<25 Mbps) | 5G NSA (W) | 5G NSA (W) | 5G NSA (W) | 5G NSA (W) | 5G NSA (W) | 5G NSA (W) | 5G NSA (W) | 5G NSA (W) | 5G NSA (W) |
| Pakistan | S2 (<200 Mbps) | 5G NSA (W) | 5G NSA (W) | 5G NSA (W) | 5G NSA (W) | 5G NSA (W) | 5G NSA (W) | 5G NSA (W) | 5G NSA (W) | 5G NSA (W) |
| Pakistan | S3 (<400 Mbps) | 5G NSA (W) | 5G NSA (W) | 5G NSA (W) | 5G NSA (W) | 5G NSA (W) | 5G NSA (W) | 5G NSA (W) | 5G NSA (W) | 5G NSA (W) |
| Albania | S1 (<25 Mbps) | 4G (W) | 4G (W) | 4G (W) | 4G (W) | 4G (W) | 4G (W) | 4G (W) | 4G (W) | 4G (W) |
| Albania | S2 (<200 Mbps) | 4G (W) | 4G (W) | 4G (W) | 4G (W) | 4G (W) | 4G (W) | 4G (W) | 4G (W) | 4G (W) |
| Albania | S3 (<400 Mbps) | 4G (W) | 4G (W) | 4G (W) | 4G (W) | 4G (W) | 4G (W) | 4G (W) | 4G (W) | 4G (W) |
| Peru | S1 (<25 Mbps) | 4G (W) | 4G (W) | 4G (W) | 4G (W) | 4G (W) | 4G (W) | 4G (W) | 4G (W) | 4G (W) |
| Peru | S2 (<200 Mbps) | 4G (W) | 4G (W) | 4G (W) | 4G (W) | 4G (W) | 4G (W) | 4G (W) | 4G (W) | 4G (W) |
| Peru | S3 (<400 Mbps) | 4G (W) | 4G (W) | 4G (W) | 4G (W) | 4G (W) | 4G (W) | 4G (W) | 4G (W) | 4G (W) |
| Mexico | S1 (<25 Mbps) | 4G (W) | 4G (W) | 4G (W) | 4G (W) | 4G (W) | 4G (W) | 4G (W) | 4G (W) | 4G (W) |
| Mexico | S2 (<200 Mbps) | 4G (W) | 4G (W) | 4G (W) | 4G (W) | 4G (W) | 4G (W) | 4G (W) | 4G (W) | 4G (W) |
| Mexico | S3 (<400 Mbps) | 4G (W) | 4G (W) | 4G (W) | 4G (W) | 4G (W) | 4G (W) | 4G (W) | 4G (W) | 4G (W) |

Second, of the four different technological options considered in the modeling, only two ever emerge as least cost for the developing countries considered. These are either 4G (W) (identified in blue font in Table 3) or 5G NSA (identified in black font in Table 3). Notably, both technological options rely on a wireless backhaul, with the use of fiber optic to reach universal access never resulting to be more cost-effective. In fact, the choice of fiber versus wireless backhaul, turns out to be a larger cost-driver than the choice of 5G over 4G. The findings also illustrate that 5G SA never proves to be a cost-effective technology for meeting universal access goals in any of the country clusters.

Third, the findings suggest that technological leapfrogging to 5G NSA can be an advisable strategy for countries with under-developed telecom infrastructure. While 5G is more cost efficient (per bit of data transferred) than previous generations of technology, the real choice faced by policy makers is between completing a partial 4G roll-out or beginning on an entirely new 5G roll-out. Hence, the existing level of 4G roll-out has a large impact on technology choice (see Table 5). This is the reason why it is cheaper overall to complete 4G deployment in countries like Albania, Peru and Mexico, where 4G coverage is already relatively high (85-95%). Whereas in countries like Senegal, Kenya and Pakistan, where current 4G coverage is much lower (30-65%), it is more attractive to deploy 5G NSA as the means of reaching universal service.

### 4.2 Impact of infrastructure sharing

As discussed, policy makers can significantly influence the cost of reaching universal access to broadband by creating a regulatory environment that supports different degrees of infrastructure sharing, which range from sharing passive infrastructure (e.g. site compounds, towers etc.), to active sharing of backhaul, through to promoting a single Shared Rural Network in areas where provision of multiple infrastructures would not be viable. For brevity, the social costs of these different forms of infrastructure sharing are reported in Table 6, for the illustrative case of 5G NSA. Complete results for all technologies are available in the Supplementary Materials Section 3 and give broadly similar results.

Table 6 Social cost results for infrastructure sharing for 5G NSA (W) technology (NPV 2020-2030)

Infrastructure Sharing Results by Country

| Scenario | Strategy | Metric | C1 Malawi | Uganda | C2 Senegal | Kenya | C3 Pakistan | C4 Albania | C5 Peru | C6 Mexico |
|---|---|---|---|---|---|---|---|---|---|---|
| S1 (<25 Mbps) | Baseline | Social Cost ($Bn) | 1.5 | 1.4 | 0.87 | 2.9 | 12 | 0.19 | 6.2 | 35 |
| S1 (<25 Mbps) | Passive | Social Cost ($Bn) | 1.3 | 1.3 | 0.74 | 2.5 | 9.2 | 0.18 | 5.1 | 30 |
| S1 (<25 Mbps) | Active | Social Cost ($Bn) | 0.79 | 0.8 | 0.39 | 1.2 | 3.5 | 0.11 | 2 | 16 |
| S1 (<25 Mbps) | SRN | Social Cost ($Bn) | 0.39 | 0.48 | 0.29 | 0.98 | 2.9 | 0.064 | 1.6 | 12 |
| S2 (<200 Mbps) | Baseline | Social Cost ($Bn) | 2.1 | 3 | 1.6 | 4.9 | 53 | 0.25 | 8.6 | 48 |
| S2 (<200 Mbps) | Passive | Social Cost ($Bn) | 1.8 | 2.1 | 1.3 | 3.9 | 34 | 0.22 | 6.9 | 39 |
| S2 (<200 Mbps) | Active | Social Cost ($Bn) | 0.9 | 1.2 | 0.64 | 1.9 | 9 | 0.13 | 2.6 | 21 |
| S2 (<200 Mbps) | SRN | Social Cost ($Bn) | 0.5 | 0.87 | 0.54 | 1.6 | 8.4 | 0.083 | 2.2 | 16 |
| S3 (<400 Mbps) | Baseline | Social Cost ($Bn) | 2.8 | 4.1 | 1.9 | 6.1 | 82 | 0.32 | 11 | 54 |
| S3 (<400 Mbps) | Passive | Social Cost ($Bn) | 2.3 | 2.9 | 1.5 | 4.9 | 54 | 0.29 | 8.5 | 43 |
| S3 (<400 Mbps) | Active | Social Cost ($Bn) | 1.1 | 1.4 | 0.72 | 2.3 | 12 | 0.16 | 3.2 | 22 |
| S3 (<400 Mbps) | SRN | Social Cost ($Bn) | 0.6 | 1 | 0.62 | 2 | 12 | 0.11 | 2.7 | 18 |

[1] Technology Strategy: 5G NSA with Wireless Backhaul.
[2] Spectrum Pricing Strategy: Baseline.
[3] Taxation Strategy: Baseline.
[4] Results rounded to 2 s.f.

The extent of infrastructure sharing is inversely related to the social cost of reaching universal access (Table 5). The savings provided by passive infrastructure sharing are not insignificant at 15-20% but are dwarfed by much higher savings of around 60% available through active infrastructure sharing, and 70% that result from adopting an Shared Rural Network. In general, the magnitude of the savings increases with the level of capacity demanded (S1-S3), but only by a few percentage points.

However, there are important caveats. Historically, competition among MNOs has led to falling costs and facilitated rapid market expansion, at least in viable areas. While infrastructure sharing may provide substantial cost savings, by removing competitive pressures between MNOs, these savings could be eroded over time due to mounting inefficiency. Going forward, a balancing act is needed, focusing on blended strategies, which preserve market dynamics in urban and suburban areas (where networks are often capacity-constrained), but promote infrastructure sharing in rural and remote locations (where networks are often coverage-constrained). In extreme situations, where no more than a single network provider is commercially viable, competition may cease to be feasible altogether, and an approach such as the Shared Rural Network may be appropriate.

4.3    Impact of spectrum pricing

Spectrum pricing carries significant weight in the cost structure for broadband infrastructure, ranging from 2-33 percent of the total industry cost across countries (Figure 5). Policy makers' decisions about how aggressively to price spectrum resources, thus materially affects the cost of reaching universal access. To explore the sensitivity of costs to the lowering or raising of spectrum prices relative to baseline levels, Table 7 reports the private, government and social costs for S2 based on 5G NSA technology strategy. Complete results for all technologies and capacity levels are available in the Supplementary Materials Section 3 and give broadly similar results.

Figure 5 Breakdown of private costs

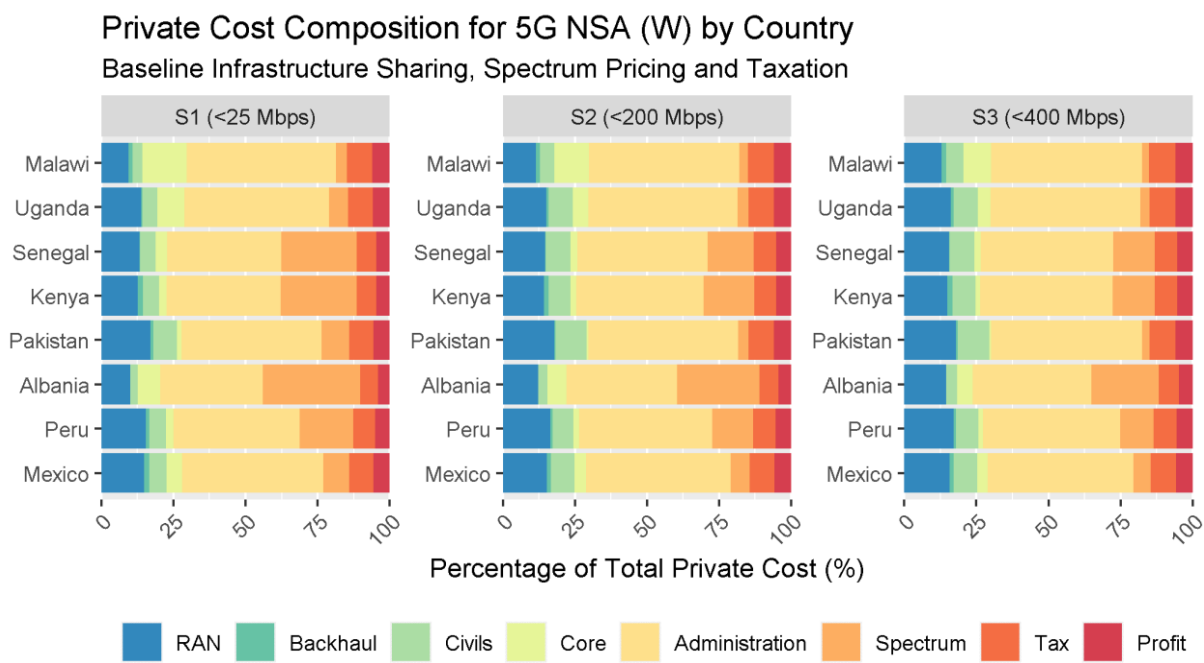

From a private sector standpoint, the difference between low and high spectrum cost scenarios can increase the costs of rolling out broadband infrastructure by between 10-30% in most countries. This factor is much higher for the case of Albania where the impact on the private cost is over 60%, reflecting particularly high baseline levels of spectrum pricing in the country.

Table 7 Breakdown of social cost results for spectrum pricing based on 5G NSA (W) and S2

Spectrum Pricing Results by Country

|  |  | C1 |  | C2 |  | C3 | C4 | C5 | C6 |
|---|---|---|---|---|---|---|---|---|---|
| Strategy | Metric | Malawi | Uganda | Senegal | Kenya | Pakistan | Albania | Peru | Mexico |
| Low Prices (-75%) | Private Cost ($Bn) | 1.70 | 2.90 | 1.90 | 5.70 | 38.00 | 0.30 | 9.90 | 54.00 |
| High Prices (+100%) | Private Cost ($Bn) | 1.70 | 3.10 | 2.50 | 7.70 | 40.00 | 0.49 | 13.00 | 61.00 |
| **Low-High Ratio (%)** | **Low-High Ratio (%)** | **100.00** | **107.00** | **132.00** | **135.00** | **105.00** | **163.00** | **131.00** | **113.00** |
| Low Prices (-75%) | Govt Cost ($Bn) | 0.45 | -0.02 | -0.25 | -0.79 | 15.00 | -0.05 | -1.30 | -5.90 |
| High Prices (+100%) | Govt Cost ($Bn) | 0.45 | -0.02 | -0.85 | -2.80 | 15.00 | -0.24 | -4.00 | -13.00 |
| **Low-High Ratio (%)** | **Low-High Ratio (%)** | **100.00** | **100.00** | **340.00** | **354.00** | **100.00** | **462.00** | **308.00** | **220.00** |
| Low Prices (-75%) | Social Cost ($Bn) | 2.10 | 2.90 | 1.60 | 4.90 | 52.00 | 0.25 | 8.60 | 48.00 |
| High Prices (+100%) | Social Cost ($Bn) | 2.20 | 3.10 | 1.60 | 4.90 | 55.00 | 0.25 | 8.60 | 48.00 |
| **Low-High Ratio (%)** | **Low-High Ratio (%)** | **105.00** | **107.00** | **100.00** | **100.00** | **106.00** | **100.00** | **100.00** | **100.00** |

[1] Scenario: S2 (<200 Mbps).
[2] Technology Strategy: 5G NSA with Wireless Backhaul.
[3] Infrastructure Sharing Strategy: Baseline.
[4] Taxation Strategy: Baseline.
[5] Results rounded to 2 d.p.

From a government standpoint, the countries divide into two groups. In a first group of countries (comprising Albania, Kenya, Mexico, Peru and Senegal), universal access to broadband is commercially viable without recourse to government subsidy. This makes the broadband sector a net contributor to the public purse through fiscal flows, so that the higher level of spectrum prices dramatically increases government revenues from the sector by as much as 200-500%. In a second group of countries (comprising Malawi, Pakistan and Uganda), universal access to broadband is not commercially viable based on user cross-subsidization alone, requiring some commitment of government subsidy. In Malawi and Pakistan (but not quite Uganda), the universal service subsidy requirement of the sector entirely offsets the fiscal revenues it generates through spectrum pricing and taxation, making broadband infrastructure a net recipient of public funds. In such challenging environments, every dollar gained from spectrum fees translates to a dollar more of required public subsidy, leaving the government's net fiscal position completely unchanged. Thus, any attempts to raise government revenues through spectrum pricing simply generates an off-setting subsidy requirement, with no resulting net impact on government finances.

Nevertheless, it is important to emphasize that any reduction of spectrum pricing would need to be accompanied by a strong regulatory regime to ensure certain objectives are achieved. For example, spectrum pricing regimes need to have coverage obligations attached to avoid private MNOs cashing in on tax breaks via extraction of excess profits from highly viable areas. Achieving total coverage involves user cross-subsidization from profitable (predominantly urban) areas to (predominantly rural and remote) locations that are not commercially viable to cover, requiring incentivization via regulatory instruments such as coverage obligations.

4.4   Subsidy requirement

In principle, it is possible to combine the cost reducing measures explored above by simultaneously permitting infrastructure sharing, while keeping spectrum pricing and taxes at relatively low levels. For the illustrative case of 5G NSA (W), the overall impact of adopting the full range of cost minimization measures is to bring costs down by as much as 75% relative to the baseline, with Pakistan being a good example (Table 8a).

In most countries considered, universal access to broadband is commercially viable for 5G NSA technology. However, for countries where this is not the case (such as Malawi, Pakistan and Uganda), such cost minimization measures can make all the difference between universal service being commercially viable or not. For example, in Pakistan in S2 (<200 Mbps), commercially viable coverage increases from below 10% to 100% when cost minimization measures are adopted (Table 8b), leading to substantial savings in the requirement for government subsidy, which would otherwise have been as high as $5 billion in NPV terms (Table 8c).

Table 8 Impact of Cost Minimization on Cost, Viability and Subsidy Requirement

(A) Social Cost of Universal Access NPV 2020-2030 by Country

| Scenario | Strategy | C1 Malawi | C1 Uganda | C2 Senegal | C2 Kenya | C3 Pakistan | C4 Albania | C5 Peru | C6 Mexico |
|---|---|---|---|---|---|---|---|---|---|
| S1 (<25 Mbps) | Baseline ($Bn) | 1.5 | 1.4 | 0.9 | 2.9 | 11.8 | 0.2 | 6.2 | 34.7 |
| S1 (<25 Mbps) | Lowest ($Bn) | 0.4 | 0.5 | 0.3 | 1 | 2.9 | 0.1 | 1.6 | 11.6 |
| S2 (<200 Mbps) | Baseline ($Bn) | 2.1 | 3 | 1.6 | 4.9 | 53.2 | 0.2 | 8.6 | 48.5 |
| S2 (<200 Mbps) | Lowest ($Bn) | 0.5 | 0.9 | 0.5 | 1.6 | 8.4 | 0.1 | 2.2 | 16.2 |
| S3 (<400 Mbps) | Baseline ($Bn) | 2.8 | 4.1 | 1.9 | 6.1 | 81.6 | 0.3 | 10.8 | 53.7 |
| S3 (<400 Mbps) | Lowest ($Bn) | 0.6 | 1 | 0.6 | 2 | 11.8 | 0.1 | 2.7 | 17.9 |

[1] Technology Strategy: 5G NSA with Wireless Backhaul.
[2] Infrastructure Sharing Strategy: Baseline.
[3] Taxation Strategy: Baseline.
[4] Results rounded to 1 d.p.

(B) Commercially Viable Population Coverage by Country

| Scenario | Strategy | C1 Malawi | C1 Uganda | C2 Senegal | C2 Kenya | C3 Pakistan | C4 Albania | C5 Peru | C6 Mexico |
|---|---|---|---|---|---|---|---|---|---|
| S1 (<25 Mbps) | Baseline (%) | 80 | 100 | 100 | 100 | 100 | 100 | 100 | 100 |
| S1 (<25 Mbps) | Lowest (%) | 100 | 100 | 100 | 100 | 100 | 100 | 100 | 100 |
| S2 (<200 Mbps) | Baseline (%) | 40 | 90 | 100 | 100 | 0 | 100 | 100 | 100 |
| S2 (<200 Mbps) | Lowest (%) | 100 | 100 | 100 | 100 | 100 | 100 | 100 | 100 |
| S3 (<400 Mbps) | Baseline (%) | 20 | 50 | 100 | 100 | 0 | 100 | 100 | 100 |
| S3 (<400 Mbps) | Lowest (%) | 100 | 100 | 100 | 100 | 100 | 100 | 100 | 100 |

[1] Technology Strategy: 5G NSA with Wireless Backhaul.
[2] Infrastructure Sharing Strategy: Baseline.
[3] Taxation Strategy: Baseline.
[4] Results rounded to 1 d.p.

(C) Government Subsidy to Reach Universal Access NPV 2020-2030 by Country

| Scenario | Strategy | C1 Malawi | C1 Uganda | C2 Senegal | C2 Kenya | C3 Pakistan | C4 Albania | C5 Peru | C6 Mexico |
|---|---|---|---|---|---|---|---|---|---|
| S1 (<25 Mbps) | Baseline ($Bn) | 0.04 | 0 | 0 | 0 | 0 | 0 | 0 | 0 |
| S1 (<25 Mbps) | Lowest ($Bn) | 0 | 0 | 0 | 0 | 0 | 0 | 0 | 0 |
| S2 (<200 Mbps) | Baseline ($Bn) | 0.09 | 0.12 | 0 | 0 | 2.77 | 0 | 0 | 0 |
| S2 (<200 Mbps) | Lowest ($Bn) | 0 | 0 | 0 | 0 | 0 | 0 | 0 | 0 |
| S3 (<400 Mbps) | Baseline ($Bn) | 0.16 | 0.28 | 0 | 0 | 4.94 | 0 | 0 | 0 |
| S3 (<400 Mbps) | Lowest ($Bn) | 0 | 0 | 0 | 0 | 0 | 0 | 0 | 0 |

[1] Technology Strategy: 5G NSA with Wireless Backhaul.
[2] Infrastructure Sharing Strategy: Baseline.
[3] Taxation Strategy: Baseline.
[4] Results rounded to 1 d.p.

4.5     Aggregation to global level

Results have so far been presented at the level of the individual countries selected to represent each of the six country clusters (recall Figure 1). These country level results can be aggregated to the global level, using the mean cost per capita within each country cluster, and thereby providing an estimate for the cost of reaching universal access to broadband across all emerging markets. Hence, Table 9 reports the total cost for universal service with each technology across the developing world, contrasting the baseline estimate with a lowest cost scenario where full infrastructure sharing is permitted, while spectrum prices and taxation are kept at their lowest level. The results are reported both in absolute dollar terms for the NPV over the period 2020-2030, as well as on an annualized basis as a share of GDP over the same period.

The first point to note is the dramatic impact of combining cost reduction measures. Across all possible technologies and scenarios for capacity demanded per user, the impact of these policy measures is to reduce the cost of meeting universal access to broadband by approximately three quarters. Focusing on the 5G NSA (W) technology option, which is lowest cost for most countries, the NPV cost of meeting universal access ranges between $0.94-1.7 trillion depending on the capacity level (S1 versus S3), but drops to the range $0.27-0.49 trillion when the full range of cost minimization measures are introduced. In GDP terms, the same costs translate into 0.31-0.57% of GDP for the baseline scenario, and 0.18-0.32% of GDP for the minimum cost scenario. Indeed, once cost minimization measures are adopted, even 5G SA can be deployed for under 1% of GDP, which would otherwise absorb 0.09-0.16% of GDP.

Table 9 Total costs for all technologies across the developing world

Technology Cost Results for the Developing World

| Scenario | Strategy | Total Cost | | 10-Year GDP Share | |
|---|---|---|---|---|---|
| | | Baseline (US$Tn) | Lowest (US$Tn) | Baseline (GDP%) | Lowest (GDP%) |
| S1 (<25 Mbps) | 4G (W) | 0.52 | 0.14 | 0.17 | 0.046 |
| S1 (<25 Mbps) | 4G (FB) | 0.72 | 0.19 | 0.24 | 0.065 |
| S1 (<25 Mbps) | 5G NSA (W) | 0.94 | 0.27 | 0.31 | 0.091 |
| S1 (<25 Mbps) | 5G SA (FB) | 2 | 0.58 | 0.67 | 0.19 |
| S2 (<200 Mbps) | 4G (W) | 1.1 | 0.28 | 0.35 | 0.093 |
| S2 (<200 Mbps) | 4G (FB) | 1.4 | 0.38 | 0.47 | 0.13 |
| S2 (<200 Mbps) | 5G NSA (W) | 1.4 | 0.41 | 0.48 | 0.14 |
| S2 (<200 Mbps) | 5G SA (FB) | 3.8 | 1 | 1.3 | 0.35 |
| S3 (<400 Mbps) | 4G (W) | 1.6 | 0.43 | 0.54 | 0.14 |
| S3 (<400 Mbps) | 4G (FB) | 2.1 | 0.55 | 0.69 | 0.18 |
| S3 (<400 Mbps) | 5G NSA (W) | 1.7 | 0.49 | 0.57 | 0.16 |
| S3 (<400 Mbps) | 5G SA (FB) | 5.2 | 1.4 | 1.7 | 0.46 |

[1] Infrastructure Sharing Strategy: Baseline.
[2] Spectrum Pricing Strategy: Baseline.
[3] Taxation Strategy: Baseline.
[4] Results rounded to 2 s.f.

Given the widely varying economic circumstances across the developing world, it is important to examine how the economic burden of reaching universal access to broadband varies across country income groupings (see Table 10). Generally, the annualized GDP share required to reach universal service is inversely related to country income level. Thus, the social cost of universal broadband access as a percentage of GDP is much higher in low-income countries (approximately 0.93-4.5%), compared to upper-middle income countries (0.11-0.93%) (for 5G NSA using wireless backhaul), with Figure 6 providing a graphical example on a country-by-country level. Considering that aggregate spending on all aspects of infrastructure in developing countries – including transport, energy and water as well as telecommunications – has recently been estimated at around 3% of GDP outside of China (Fay et al., 2019), such figures look implausibly large. However, costs can be reduced considerably if a combination of lowest cost strategies is implemented. For example, for the high capacity scenario (S3) favored by the UN Broadband Commission, the social cost of a 5G NSA (W) approach based on full adoption of cost minimization measures can be reduced to 0.54%, 0.24% and 0.14% of GDP for low, lower-middle and upper-middle income countries respectively. Such figures start to look somewhat more affordable within the overall spending envelope on infrastructure.

Table 10 Cost as a share of GDP broken down by income group

Technology Cost Results for the Developing World

| Scenario | Strategy | Low Income (10-Year GDP%) | | Lower Middle Income (10-Year GDP%) | | Upper Middle Income (10-Year GDP%) | |
|---|---|---|---|---|---|---|---|
| | | Baseline LIC | Lowest LIC | Baseline LMIC | Lowest LMIC | Baseline UMIC | Lowest UMIC |
| S1 (<25 Mbps) | 4G (W) | 0.93 | 0.28 | 0.39 | 0.097 | 0.11 | 0.03 |
| S1 (<25 Mbps) | 4G (FB) | 1.4 | 0.44 | 0.56 | 0.14 | 0.15 | 0.041 |
| S1 (<25 Mbps) | 5G NSA (W) | 0.96 | 0.29 | 0.32 | 0.083 | 0.3 | 0.091 |
| S1 (<25 Mbps) | 5G SA (FB) | 2.4 | 0.71 | 0.86 | 0.22 | 0.6 | 0.18 |
| S2 (<200 Mbps) | 4G (W) | 1.7 | 0.53 | 0.96 | 0.24 | 0.19 | 0.053 |
| S2 (<200 Mbps) | 4G (FB) | 2.4 | 0.74 | 1.3 | 0.33 | 0.24 | 0.069 |
| S2 (<200 Mbps) | 5G NSA (W) | 1.5 | 0.45 | 0.74 | 0.18 | 0.4 | 0.12 |
| S2 (<200 Mbps) | 5G SA (FB) | 3.6 | 1.1 | 3 | 0.72 | 0.81 | 0.25 |
| S3 (<400 Mbps) | 4G (W) | 2.8 | 0.83 | 1.7 | 0.41 | 0.24 | 0.069 |
| S3 (<400 Mbps) | 4G (FB) | 3.6 | 1.1 | 2.1 | 0.52 | 0.31 | 0.089 |
| S3 (<400 Mbps) | 5G NSA (W) | 1.8 | 0.54 | 1 | 0.24 | 0.45 | 0.14 |
| S3 (<400 Mbps) | 5G SA (FB) | 4.5 | 1.3 | 5 | 1.2 | 0.93 | 0.28 |

[1] Infrastructure Sharing Strategy: Baseline.
[2] Spectrum Pricing Strategy: Baseline.
[3] Taxation Strategy: Baseline.
[4] Results rounded to 2 s.f.

Figure 6 Example country by country investment for 5G NSA (Wireless)

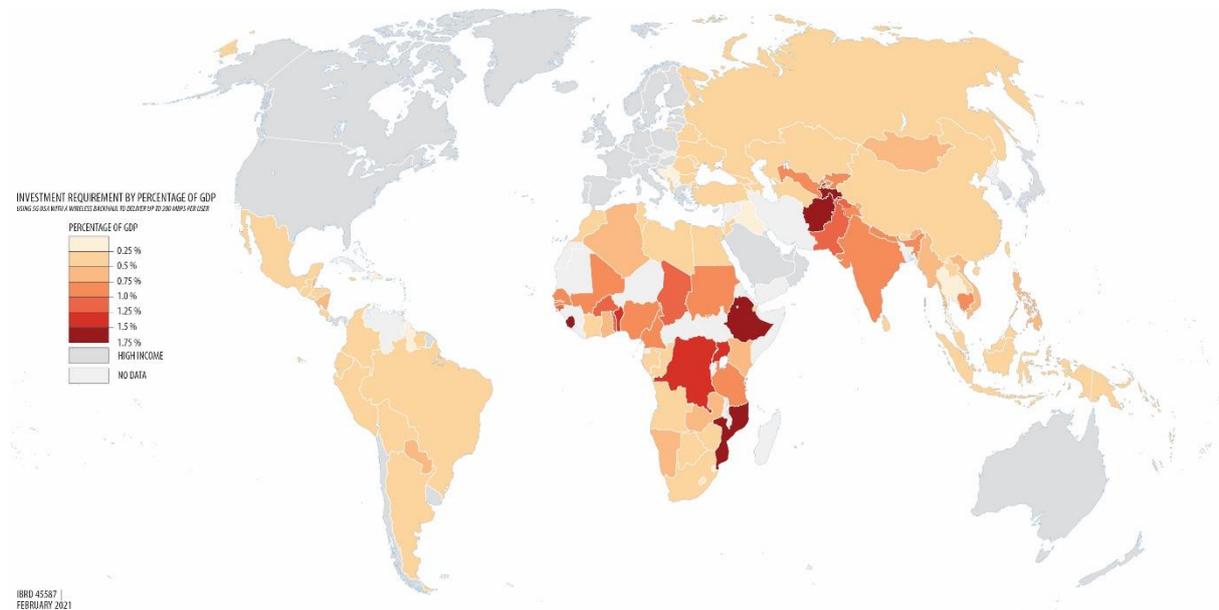

5. Conclusions

In this analysis, a scenario-based approach was adopted to model the deployment of broadband infrastructure in 8 developing countries across 6 different country clusters. The results were then aggregated across all developing countries to provide insight into the social costs of meeting goals of universal access to broadband, and how these are affected by government policy choices regarding preferred technology, the extent of infrastructure sharing, and the level of spectrum pricing and taxation.

The baseline social cost of broadband universal service to all developing countries over the next decade ranged from $0.52 trillion for 4G to $0.94 trillion using 5G NSA (10-Year GDP shares of 0.17% and 0.31% respectively). This is based on meeting minimum capacity requirements for urban (25 Mbps), suburban (10 Mbps) and rural (2 Mbps) users. The cost of meeting the UN Broadband Commission target is estimated at $1.6 trillion using 4G and $1.7 trillion using 5G NSA, equating to approximately 0.5% and 0.6% of annual GDP for the developing world over the next decade. This is based on meeting minimum capacity requirements for urban (400 Mbps), suburban (100 Mbps) and rural (10 Mbps) users.

Governments should consider carefully the choice between 4G or 5G NSA technology as the basis for their universal service strategies. The study finds that 5G NSA, can be the lowest cost technology choice in many developing country settings thanks to the added spectral efficiency it can provide, leading to a reduction in the number of required sites. This is the case in countries where 4G networks remain incomplete, reaching no more than 50-60% of the population (such as Kenya, Pakistan and Senegal), and indicates the potential for technological leapfrogging. On the other hand, in countries that have already made significant progress in extending 4G coverage to 80-90% of the population (such as Albania, Mexico, Peru), it proves more cost-effective to reach universal access by completing the roll-out of 4G networks. Essentially, what matters is whether a country is mostly capacity-constrained (requiring lots of spectrum and cells to meet large traffic demand) or mostly coverage-constrained (requiring lots of cells to thinly cover very large geographic areas). The findings suggest that 5G deployment (specifically 5G NSA with wireless backhaul) is much more viable in capacity-constrained countries such as Pakistan, in contrast with coverage-constrained countries such as Malawi or Uganda, where 4G is competitive at providing low user capacities.

Whether 4G or 5G NSA technology is selected, in both cases wireless backhaul proves to be much lower cost than reliance on fiber optic to reach such outlying areas. However, this does not contradict the essential importance of fiber backhaul to support both 4G and 5G technologies (as well as future generations) in high demand areas.

Government policy choices can dramatically reduce the cost of reaching universal broadband access, and thereby improve commercial viability avoiding (or at least reducing) the need for public subsidy. Across the two most plausible technological choices, 4G (W) and 5G NSA, the creation of a regulatory environment supportive of infrastructure sharing is predicted to reduce the social cost of reaching universal broadband access by ~15-72%, with a Shared Rural Network providing the largest cost reduction. If, at the same time, spectrum pricing is kept at about half of historic levels, overall savings relative to the baseline costs reported above could be further reduced by up to 75% of the required investment. While raising spectrum revenue can be alluring for governments, the results demonstrate how this increases the social cost of broadband universal service. Moreover, in less developed contexts, where universal service may not be commercially viable and government subsidy is required to close the viability gap, raising spectrum pricing is found to have no net impact on government revenues as it simply translates, dollar for dollar, into a larger subsidy requirement for the sector.

Future research should seek to apply the model to understand how many MNOs can viably be supported in different locations, ensuring competition benefits can be maximized. By identifying those places where a single network cannot be viably supported, these will be prime locations to explore implementing more radical options such as a Shared Rural Network. Further development of the *pytal* software should focus on better defining both supply-side and demand-side taxation, in order to provide a more realistic reflection of the fiscal burden affecting operators in unviable locations (given the analysis here focused only on supply-side decisions). Additionally, a range of new Low Earth Orbit satellite constellations are being deployed to provide broadband in rural and remote areas, therefore future research should consider the comparative cost implications of using these technologies. Finally, as this analysis focused on cost-minimization, future research should look to introduce the potential net benefits from roll-out which were beyond the scope of this paper.

Acknowledgements

The authors have no competing interests. In the process of undertaking the research we would like to thank the governments of Kenya, Senegal and Albania for providing cellular site data. We also valued review meetings with (i) telecommunications regulators in Malawi, Uganda, Kenya, Senegal, Pakistan, Albania, Peru and Mexico and (ii) mobile network operators including Entel (Peru), AT&T (Mexico), Telenor (Pakistan), Sonatel (Senegal) and TNM (Malawi). We also appreciated the valuable feedback from the World Economic Forum and the GSMA. Additionally, the authors thank various individuals for providing specific expertise on parts of the paper, including Peter Curnow-Ford, Adnan Shahid, Andy Sutton and Tom Russell, and Andoria Indah Purwaningtyas for operational support. This work was supported by (i) UKRI grant EP/N017064/1 and an EPSRC Impact Accelerator Award, and (ii) the World Bank 5G Flagship.

**Edward Oughton** received the M.Phil. and Ph.D. degrees from Clare College, at the University of Cambridge, U.K., in 2010 and 2015, respectively. He later held research positions at both Cambridge and Oxford. He is currently an Assistant Professor in the College of Science at George Mason University, Fairfax, VA, USA, developing open-source research software to analyze digital infrastructure deployment strategies. He received the Pacific Telecommunication Council Young Scholars Award in 2019, Best Paper Award 2019 from the Society of Risk Analysis, and the TPRC48 Benton Early Career Award 2021.

**Niccolò Comini** is an Economist in the Digital Development Global Practice of the Infrastructure Vice Presidency at the World Bank Group focusing on ICT competition and regulatory issues in emerging countries. Examples include advising governments on 5G network deployment, data infrastructure, and digital platforms. Prior to joining the World Bank, he worked at the OECD in Mexico City and Paris as a competition expert. Niccolò is fluent in Italian, English and Spanish. He holds a MSc in Competition and Market Regulation from the Barcelona Graduate School of Economics and a M.A. in Law and Economics from the University of Bologna.

**Vivien Foster** is the Chief Economist for the Infrastructure Vice-Presidency of the World Bank covering Digital Development, Energy & Extractives, Transport and Infrastructure Finance. During her 20 years at the World Bank she has played a variety of leadership roles. Prior to joining the World Bank, Vivien worked as a Managing Consultant of Oxford Economic Research Associates Ltd in the UK. She is a graduate of Oxford University, and also holds a Master's from Stanford University and a Doctorate from University College London, both in Economics.

**Jim Hall** is Professor of Climate and Environmental Risks in the University of Oxford and Director of Research in the School of Geography and the Environment. Before joining the University of Oxford in 2011, Prof Hall held academic positions in Newcastle University and the University of Bristol. Prof Hall is internationally recognized for his research on risk analysis and decision making under uncertainty for water resource systems, flood and coastal risk management, infrastructure systems and adaptation to climate change.

Supplementary Materials

1. Clustering

Clustering is a popular classical unsupervised machine learning method for multivariate datasets. However, the approach does require the exogenous pre-specification of the number of clusters. To aid in this process, the plotting of the Within-Group Sum of Squares (WSS) for many defined cluster combinations is required. The aim is to reduce the over variation within each group, as indicated by the WSS metric (Hothorn and Everitt, 2014). Mean centered values are used for the clustering process. Reductions in variation for different cluster sizes can be identified based on multiple runs of the algorithm, as shown in Figure S1 below.

Figure S1 WSS results plotted against the potential number of clusters

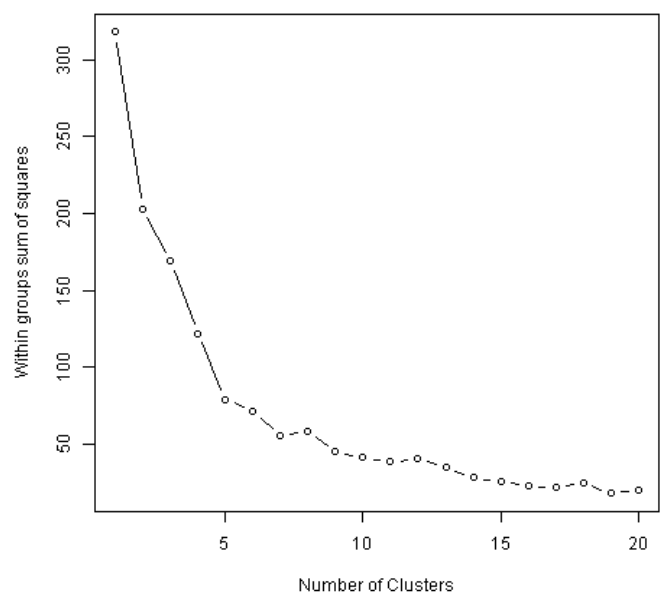

The *kmeans* function from the inbuilt R *stats* package is applied. A scaled numeric matrix is first created which enables the use of the *kmeans* function, with the code iterating over the potential number of exogenously specified clusters (referred to as 'centers' in the *kmeans* R package). Figure illustrates the reduction in variance is significant as the first few clusters are added (1-5), but subsequent gains are marginal. The choice of the number of clusters needs to be evaluated against the ease of reporting the results, thus, an even number of 6 clusters is selected for this analysis.

2. Method: The Python Telecommunications Assessment Library (*pytal*)

Such an approach allows the simulation of different decision options which may affect 4G and 5G deployment, given a range of future scenarios. Although such simulation models have already been applied in the literature for both 4G and 5G, they have rarely been rigorously applied for cross-country assessment.

The model will also be developed to reflect and incorporate the institutional organization of the sector, specifically the number of mobile operators, the potential roles of the public and private sector in network roll-out, as well as revenue and cost efficiency characteristics.

## 2.1. Estimating baseline infrastructure

Previous generations of infrastructure investment have led to many sunk costs in infrastructure assets, such as towers, backhaul and core nodes, which can fortunately be reused and/or upgraded. The challenge consequently is establishing the number and location of existing assets, particularly as this data is often limited due to commercial sensitivities. Site data is obtained from telecommunication regulators in Senegal, Kenya and Albania. For those countries where site data are not available, existing tower estimates are taken from TowerXchange (TowerXchange, 2019a, 2019b, 2018a, 2018b, 2017) by country and disaggregated to help establish the number of assets in each region (Uganda, 3,554; Malawi, 1,000; Pakistan, 43,300; Peru, 9,193 and Mexico, 29,159).

Equation (9) outlines the method used to estimate the number of towers ($Towers_i$) in the $i$th area.

$$Towers_i = Population_i \cdot \frac{Towers}{(Population \cdot (Coverage/100)} \tag{9}$$

This is estimated by multiplying the population ($Population_i$) of the $i$th area, by the number of towers per covered population, which is calculated using the number of known towers nationally ($Towers$), the total country population ($Population$) and the national population coverage by 2G GSM ($Coverage$). As an illustrative example, imagine trying to estimate the number of towers in a region with a population of 100. There are 5,000 towers nationally, a country population of 1,000,000 and a population coverage of 50%. Thus, Equation (10) would estimate there would be 1 tower in the region.

To ensure only areas with population coverage receive towers, all areas are sorted based on the population density, with the densest areas at the top of the list, and least dense at the bottom. Equation (11) is applied to each region beginning at the top of the list, to allocate the available assets to the areas of highest population density first (and hence highest traffic demand), as a rational infrastructure operator would behave.

Once all towers have been allocated, those areas without assets would be the least densely populated areas, which is logical with how market failures occur in the deployment of telecoms in rural and remote regions.

Finally, the technology type (legacy 2G/3G or 4G) is allocated to all assets. By intersecting coverage map data for each technology from the Mobile Coverage Explorer(Collins Bartholomew, 2019), with the sub-regional boundaries which have tower estimates, the number of sites by technology can be estimated.

## 2.2. Radio Access Network (RAN) design

Having determined the approximate number of existing infrastructure assets, it is then possible to establish the number of brownfield sites to upgrade, or new greenfield sites needing to be built. For a given level of data traffic in each area, we estimate the least-cost network design to meet demand given the available spectrum, spectral efficiency of the available technology and required number of cell sites.

An industry-standard cellular dimensioning methodology is used to assess network delivery using the eMBB 5G use case put forward by 3GPP release 15 specification(3GPP, 2019) (which contains by 4G and 5G spectral efficiencies for different Signal-to-Interference-Plus-Noise levels). The existing open-source python simulator for integrated modelling of 5G (pysim5G) is applied to develop a set of engineering-economic metrics for different types of network configurations (Oughton, 2019; Oughton et al., 2019b), which then allow least-cost estimation of network design for both 4G and 5G networks.

The aim of the pysim5G approach is to account for the three main factors that affect the capacity of a wireless network which include (i) the spectral efficiency of the technology, (ii) the frequency and bandwidth of available spectrum carrier bands, and (iii) the spatial reuse of available spectrum.

Table S1 details the frequency bands available to test, including maximum bandwidth, technology and duplex mode. All frequencies are treated as being Frequency Division Duplex (FDD), except 3.5 GHz which is treated as being Time Division Duplex (TDD) with an 70:30 downlink to uplink ratio. 4G bands are treated as being MIMO (2x2), and 5G bands are treated as being MIMO (4x4).

Table S1 Available spectrum

| Tech | Frequency (MHz) | Bandwidth (MHz) | | | | | | | |
|---|---|---|---|---|---|---|---|---|---|
| | | Malawi | Uganda | Senegal | Kenya | Pakistan | Albania | Peru | Mexico |
| 4G | 700 | | | | | | | 30 | 20 |
| | 800 | 10 | 10 | 10 | 10 | 10 | 10 | | |
| | 850 | | | | | | | | 10 |
| | 1700 | | | | | | | | 20 |
| | 1800 | 10 | | 10 | 10 | 10 | 10 | | |
| | 1900 | | | | | | | | 20 |
| | 2500 | | | | | | | | 10 |
| | 2600 | | 10 | | | | | 20 | |
| 5G | 700 | 10 | 10 | 10 | 10 | 10 | 10 | 15 | |
| | 2500 | | | | | | | | 20 |
| | 3500 | 50 | 50 | 50 | 50 | 50 | 50 | 50 | 50 |
| Total (MHz) | | 80 | 80 | 80 | 80 | 80 | 80 | 115 | 150 |

The capacity provided by a radio channel is dependent on a range of factors, which primarily includes carrier frequency propagation characteristics, inter-cell interference and channel bandwidth. Hence, Monte Carlo simulations are carried out using pysim5G to capture the average area capacity given the technology (4G or 5G), the amount of available spectrum (frequency and bandwidth) and the density of cells (sites per square kilometer). The stochastic component directly relates to the potential path loss between the radio antenna and receiver, which is particularly affected by distance and frequency. The parameters used in the stochastic simulations are defined in Table S2.

Table S2 Radio simulation parameters

| Simulation parameter | Unit | Value |
|---|---|---|
| Iterations | Samples per receiver (n) | 20 |
| Transmission method | - | 4G 2x2 MIMO, 5G 4x4 MIMO |
| Propagation model | - | ETSI TR 138 901 |
| Shadow fading (log normal) | dB | $(\mu, \sigma) = (0, \sigma)$ |
| Frequency reuse | Factor | 1 |
| Line of Sight | Meters | <500 |
| Indoor probability | Percent | 50 |
| Building penetration loss (log normal) | dB | $(\mu, \sigma) = (12, 8)$ |
| Transmit power | dBm | 40 |
| Transmitter antenna type | - | Directional |
| Transmitter antenna gain | dBi | 16 |
| Sectors | Sectors | 3 |
| Transmitter height | Meters | 30 |
| UE antenna gain | dBi | 0 |
| UE losses | dB | 4 |
| UE misc. losses | dB | 4 |
| UE height | Meters | 1.5 |
| Network load | Percent | 50 |

A variety of key capacity metrics are reported in Figure S2 to illustrate how distance from the cell site affects spectral efficiency, and hence potential capacity. For example, while the capacity of 700 MHz (10x10MHz bandwidth) leads to a per user speed of 100 Mbps close to the cell, this capacity drops below 5 Mbps at 5 kilometers. Hence, while the *mean* capacity is used in the modeling approach, the user experience would vary significantly depending on location within the cell area. Indeed, peak capacities could be significantly higher.

Figure S2 Capacity performance of 4G and 5G bands given distance

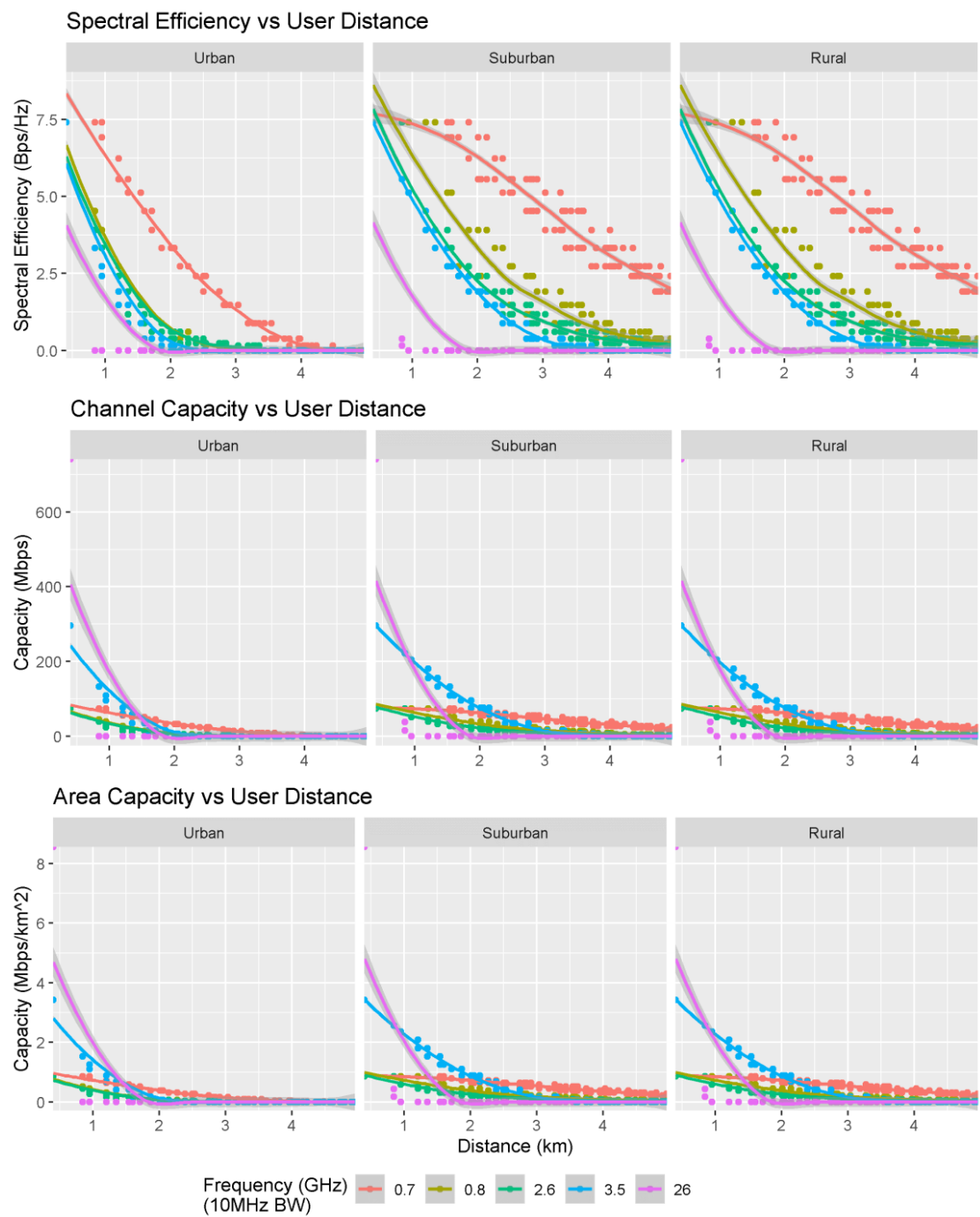

## 2.3. Backhaul/fronthaul network design

Cell sites need to connect into a local metro or core network. Depending on the network design, the fronthaul and/or backhaul cost can be one of the major cost components necessary in the delivery of a new cellular site, particularly in rural and remote locations where it can be challenging to traverse irregular terrain.

The proportion of backhaul technologies in use currently across all macro cell sites for each global region is reported in Figure S3 from GSMA (GSMA, 2019). The backhaul technology needs to be estimated for each site, which can be achieved using a sequential probability approach. Beginning in the most populated areas, the density of users best supports the use of fiber. The least populated areas are most likely to use satellite. By ranking all areas based on population density, fiber backhaul can be allocated to the first $n$ most dense sites given the stated proportion nationally, followed by the next $n$ most dense sites receiving copper fixed lines, and so on. The least dense, most remote sites are allocated satellite backhaul. In the aggregate, the backhaul technologies in place in each country match the regional profiles specified by GSMA.

Figure S3 Backhaul method by region

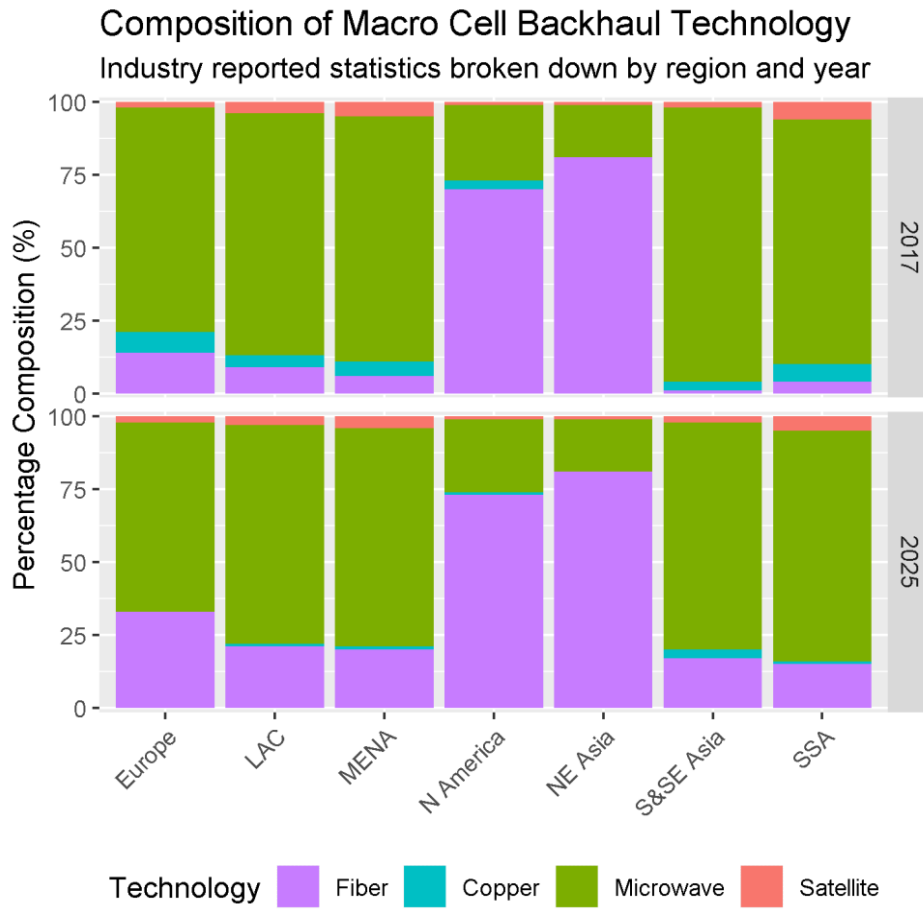

The cost of backhaul technologies often relates to the distance data needs to be transported over. In the assessment model, the mean backhaul distance ($\bar{x}$) in meters is estimated based on the density of regional or core network nodes ($NodeDensity_i$) which data must be transported to for the $i$th area given $\bar{x} = \sqrt{\frac{\frac{1}{NodeDensity_i}}{2}}$. This mean distance is used to estimate the potential cost of greenfield fiber deployment, or the cost of different types of wireless backhaul technology.

2.4. Core and regional fiber network design

The core network transports very large amounts of data traffic over long distances and connects into the global internet. We have only limited information on the core networks of many countries. Hence, in this analysis we first use all available information to assemble an existing fiber network and then design regional fiber links to provide the necessary digital infrastructure to assess 100% coverage. In all strategies tested, we extend the core network into every region, and build a local fiber network which connects all major settlements (>1000 inhabitants). This method is general enough to be applied to all countries included in the analysis. Figure S4 illustrates the processing steps, including data and parameter inputs, and the extracted data layers.

Figure S4 Core and regional fiber network estimation

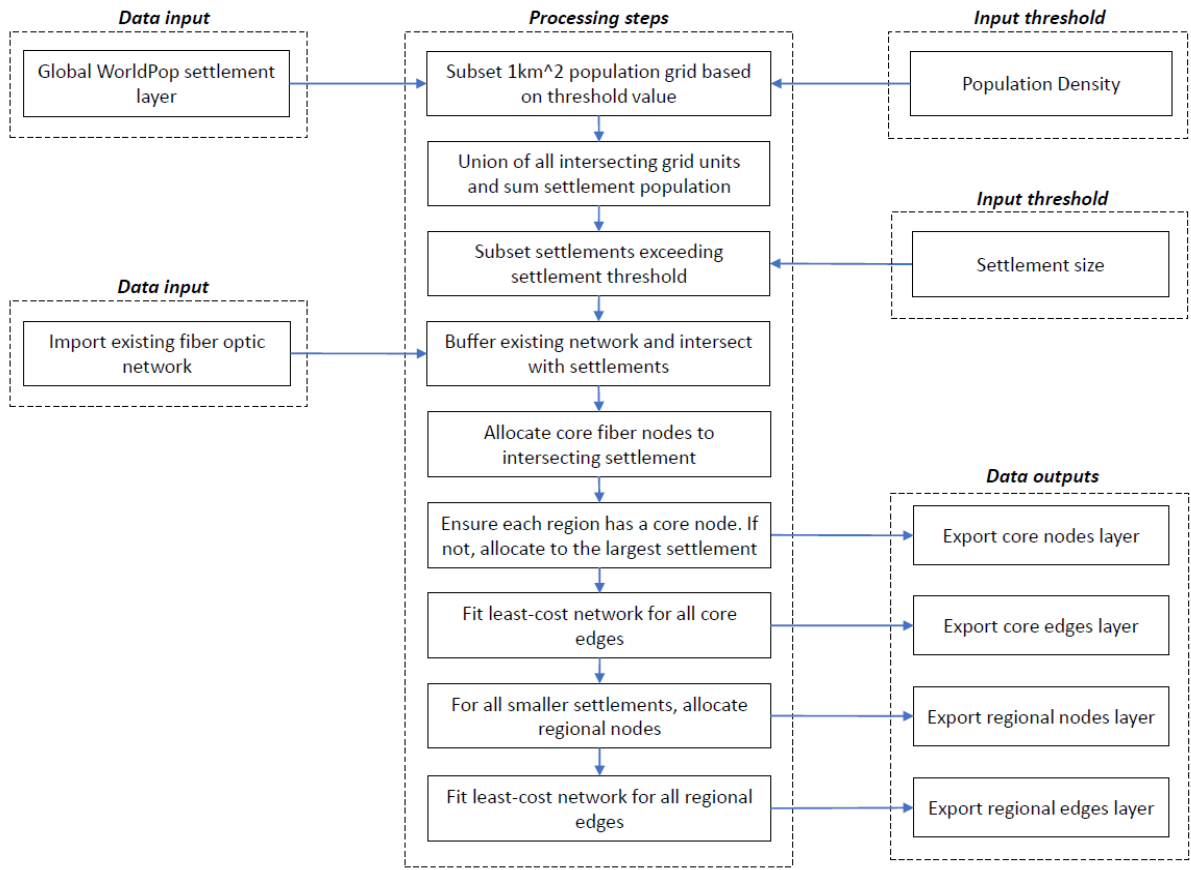

Using the WorldPop global settlement layer already used earlier in the method, high population density areas are identified which exceed a stated population density threshold ($DensityThreshold$). This threshold is set to 500 inhabitants per km$^2$ for all countries except Senegal where 200 inhabitants per km$^2$ is used (due to a smaller, less agglomerated population). A union of all intersecting high population areas is then taken, with the population being summed by settlement.

A subset is then taken of all settlements exceeding the defined settlement size ($SettlementThreshold$) which is set to 1000 inhabitants. A 2-kilometer buffer is then added to the existing fiber network, which is intersected with the settlements layer. Intersecting settlements are treated as being connected to the core network, as infrastructure operators would rationally want to connect areas where user revenues can be generated. If any regions do not have a core node, the largest settlement is selected, and new least-cost fiber connections are designed by connecting via the shortest-path to the nearest node using a Minimum Spanning Tree based on Dijkstra's algorithm. Finally, all regional nodes are linked together and back to the core, also using a Minimum Spanning Tree to obtain the least-cost design for local data transmission back to the main fiber core network.

## 2.5. Upgrade strategies

A traditional 2G or 3G site has mounted radio antennas with Mast Head Amplifiers (MHAs) fed by a coaxial feeder cable from co-located Baseband Units (BBUs) and Radio Units (RUs)(Sutton, 2018). In contrast, 4G LTE/5G NSA takes advantage of Remote Radio Units (RRUs) (Sutton, 2019) allowing BBUs to be located further away from antenna (used mainly in urban areas), but still essentially at the same site. Both 4G and 5G NSA are treated as having a Distributed Radio Access Network (D-RAN) as stated in Equation (12) which calculates the total cost the distributed RAN at the $i$th site ($RAN_{di}$).

$$RAN_{di} = Interface + FH + BBProc + GProc + Control + Alarm + Fan + Power + Cabinet \quad (12)$$

Where the estimate includes the costs of the S1-X2 or N2-N3 interface ($Interface$) (depending on 4G or 5G), fronthaul interface ($FH$), baseband processing unit ($BBProc$), general processing unit ($GProc$), control unit ($Control$), alarm unit ($Alarm$), fan/cooling unit ($Fan$), power supply including power conversion ($Power$), and the cabinet to house the equipment ($Cabinet$). This equipment is co-located on the cell site.

In contrast, a centralized Standalone 5G Cloud-RAN has a fiber fronthaul link that connects the on-site RRUs to a centralized cloud data center node which contains Virtualized Baseband Units (vBBUs) running on generic server processing hardware. Whereas the upgrading strategies for existing sites to 4G or 5G NSA are well understood from the existing literature (Oughton et al., 2019a, 2019b), the migratory path for 5G SA with C-RAN is less well-defined. Therefore, a C-RAN cost model is adopted from the literature (De Andrade et al., 2015; Shehata et al., 2017) to estimate the deployment cost of a virtualized RAN, with NFV/SDN functionality. The cost calculation is outlined in equation (13).

$$RAN_{ci} = Interface + FH + v(BBProc + GProc) + Control + Alarm + Fan + Power + v(Rack) \quad (13)$$

Many of the components are the same as in the D-RAN approach, although financial benefits accrue through resource sharing. For 5G this includes the N2-N3 interface ($Interface$), fronthaul interface ($FH$), baseband processing unit ($BBProc$), general processing unit ($GProc$), control unit ($Control$), alarm unit ($Alarm$), fan/cooling unit ($Fan$), power supply including power conversion ($Power$), and the rack to house the equipment ($Rack$). Components are shared at a splitting factor of 1:7 (N2-N3 IO interface, switch, rack, power supply). Based on a hexagonal site spacing, which is a standard design approach for cellular networks, cloud processing equipment is co-located on a central cell site which becomes designated as a 'local cloud' node as illustrated in Figure S5.

Figure S5 Cloud-RAN network hierarchy

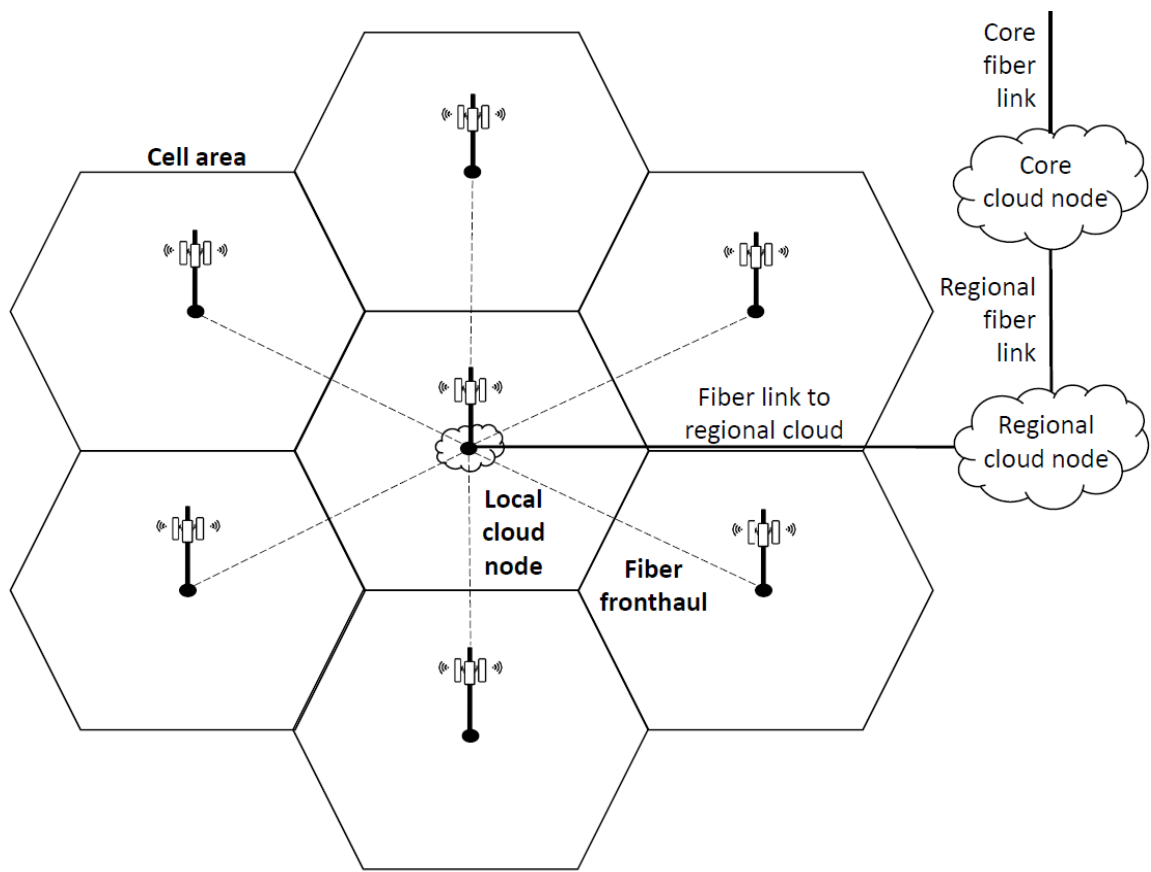

To ensure low latency, one local cloud node is deployed every 40 km² ensuring fronthaul links generally stay below 20 km. The fronthaul from other sites is connected into the local node via fiber which is based on the mean site spacing ($\bar{x}$) given the site density ($SiteDensity$) in the $i$th area per square kilometer, where $\bar{x} = \frac{\sqrt{\frac{1}{SiteDensity_i}}}{2}$. A local cloud fiber link is then required to connect to the nearest regional or core cloud node. In terms of resource sharing, the virtualization factor equates to sharing computational processing resources across cells, based on an average virtualization ratio of one blade to two cells ($v$ = 1:2) in urban, one blade to four cells ($v$ = 1:4) in suburban, and one blade to sixteen cells ($v$ = 1:16) in rural. These splitting values reflect the differences in traffic demand between dense urban and sparse rural areas.

### 2.6. Cost estimation

By examining equipment costs for 4G and 5G assessments in the literature(5G NORMA, 2016; Frias et al., 2017; METIS, 2017; Oughton et al., 2019a; Rendon Schneir et al., 2019; Wisely et al., 2018), combined with open information from regulatory LRIC models (Ofcom, 2018), capital and operational cost estimates can be determined. Following a similar method to Wisely et al.(Wisely et al., 2018), both Table S3 and Table S4 report the capex and opex costs respectively for macro cells over a 10-year period. The market for telecommunication equipment is global, therefore it is reasonable to use standardized costs in United States Dollars (USD).

Table S3 Macro cell capex estimates

| Author(s) | Equipment ($USD 1,000s) | Site build ($USD 1,000s) | Installation ($USD 1,000s) | Total ($USD 1,000s) |
|---|---|---|---|---|
| Markendahl & Mäkitalo (Markendahl and Mäkitalo, 2010) | - | - | - | 130.8 |
| Nikolikj & Janevski (Nikolikj and Janevski, 2014) | - | - | - | 130.8 |
| Johansson et al. (Johansson et al., 2004) | 54.5 | 76.3 | 32.7 | 163.5 |
| Frias & Perez (Frias and Pérez, 2012) | 36.0 | 81.8 | - | 117.7 |
| Yunas et al. (Yunas et al., 2014) | 10.9 | - | 5.5 | 16.4 |
| Smail and Weijia (Smail and Weijia, 2017) | - | - | - | 120.0 |
| Paolini (Paolini and Fili, 2012) | 36.0 | - | 48.8 | 84.8 |
| Oughton & Frias (Oughton and Frias, 2018) | 49.9 | 6.1 | 22.0 | 78.0 |
| 5G Norma (5G NORMA, 2016) | 48.6 | 22.0 | - | 70.5 |
| Mean | 39.3 | 31.0 | 18.2 | 101.4 |

Table S4 Macro cell opex estimates

| Author(s) | Site rental ($USD 1,000s) | Backhaul ($USD 1,000s) | O&M ($USD 1,000s) | Power ($USD 1,000s) | Total ($USD 1,000s) |
|---|---|---|---|---|---|
| Markendahl & Mäkitalo (Markendahl and Mäkitalo, 2010) | 8.2 | 13.1 | 8.2 | 4.4 | 33.8 |
| Johansson et al. (Johansson et al., 2004) | 10.9 | 5.5 | 3.3 | - | 19.6 |
| Frias & Perez (Frias and Pérez, 2012) | 15.3 | 10.9 | - | - | 26.2 |
| Yunas et al. (Yunas et al., 2014) | 5.5 | 2.2 | 7.6 | | 15.3 |
| Smail and Weijia (Smail and Weijia, 2017) | - | - | - | - | 30.0 |
| Paolini (Paolini and Fili, 2012) | 18.3 | 7.3 | 12.2 | - | 37.8 |
| Oughton & Frias (Oughton and Frias, 2018) | 6.1 | - | 4.9 | - | 11.0 |
| 5G Norma (5G NORMA, 2016) | 4.9 | 9.5 | 8.5 | - | 22.9 |
| Mean | 9.9 | 8.1 | 7.4 | 4.4 | 21.8 |

Equipment costs are challenging because prices are usually seen as commercially sensitive information, but the review reports that costs vary ~$10-50k, with a mean of ~$40k. These values are skewed upwards by greenfield deployment costs, rather than with brownfield upgrades where additional carriers could be added to existing multi-carrier basestations. Site building costs range from ~$6-80k, with a mean of ~$30k, and installation ranges from ~$6-50k, with a mean of $18k (also skewed upwards for greenfield sites). In terms of opex, site rental had a range of $5-20k with a mean

of $10k, operation and maintenance had a range of $3-12k with a mean of $7k, and power had too few observations to generalize. It is important to remember these costs are from High Income countries and reflect different technologies and deployment contexts. They are likely to be lower in LMIC nations due to reduced labor costs, hence require adaptation. Additionally, as it is necessary to estimate both distributed and cloud architectures, a more disaggregated approach is needed. Hence, equipment and installation costs are stated in Table S5.

Table S5 Equipment and installation costs

| Item | Unit cost ($) |
|---|---|
| Single antenna | 1,500 |
| Single remote radio unit | 4,000 |
| IO fronthaul interface | 1500 |
| Control unit | 1500 |
| Fans / cooling unit | 250 |
| Distributed power supply convertor | 250 |
| Baseband unit cabinet | 500 |
| COTS processing | 500 |
| Low latency switch | 500 |
| Rack cabinet | 500 |
| Cloud power supply converter | 1000 |
| Tower | 5,000 |
| Civil materials | 5,000 |
| Transportation | 5,000 |
| Installation | 5,000 |
| Site rental (urban) (w/security) | 9,600 |
| Site rental (suburban) (w/security) | 4,000 |
| Site rental (rural) (w/security) | 1,000 |
| Power generator and battery system(4G) | 5,000 |
| Power generator and battery system(5G) | 15,000 |
| Microwave backhaul (small) (<20km) | 30,000 |
| Microwave backhaul (medium) (20-40km) | 40,000 |
| Microwave backhaul (large) (>40km) | 80,000 |
| Fiber (urban) (per meter) | 25 |
| Fiber (suburban) (per meter) | 15 |
| Fiber (rural) (per meter) | 5 |
| Core node (4G EPC) | 75,000 |
| Core node (5G NSA) | 75,000 |
| Core cloud node (5G SA) | 250,000 |
| Core fiber link (per meter) | 4 |
| Regional node (4G EPC) | 20,000 |
| Regional node (5G NSA) | 20,000 |
| Regional cloud node (5G SA) | 100,000 |
| Regional fiber link (per meter) | 2 |
| Local node (4G EPC) | 20,000 |
| Local node (5G NSA) | 20,000 |
| Local cloud node (5G SA) | 100,000 |

2.7. Spectrum costs

Figure S6 illustrates the statistical distribution of historical auction prices paid for different types of spectrum based on the six cluster groupings identified. Prices have been adjusted for license duration to obtain the dollar cost per MHz per head of the population, hence adjusting for frequency bandwidth and market potential.

Figure S6 Spectrum costs

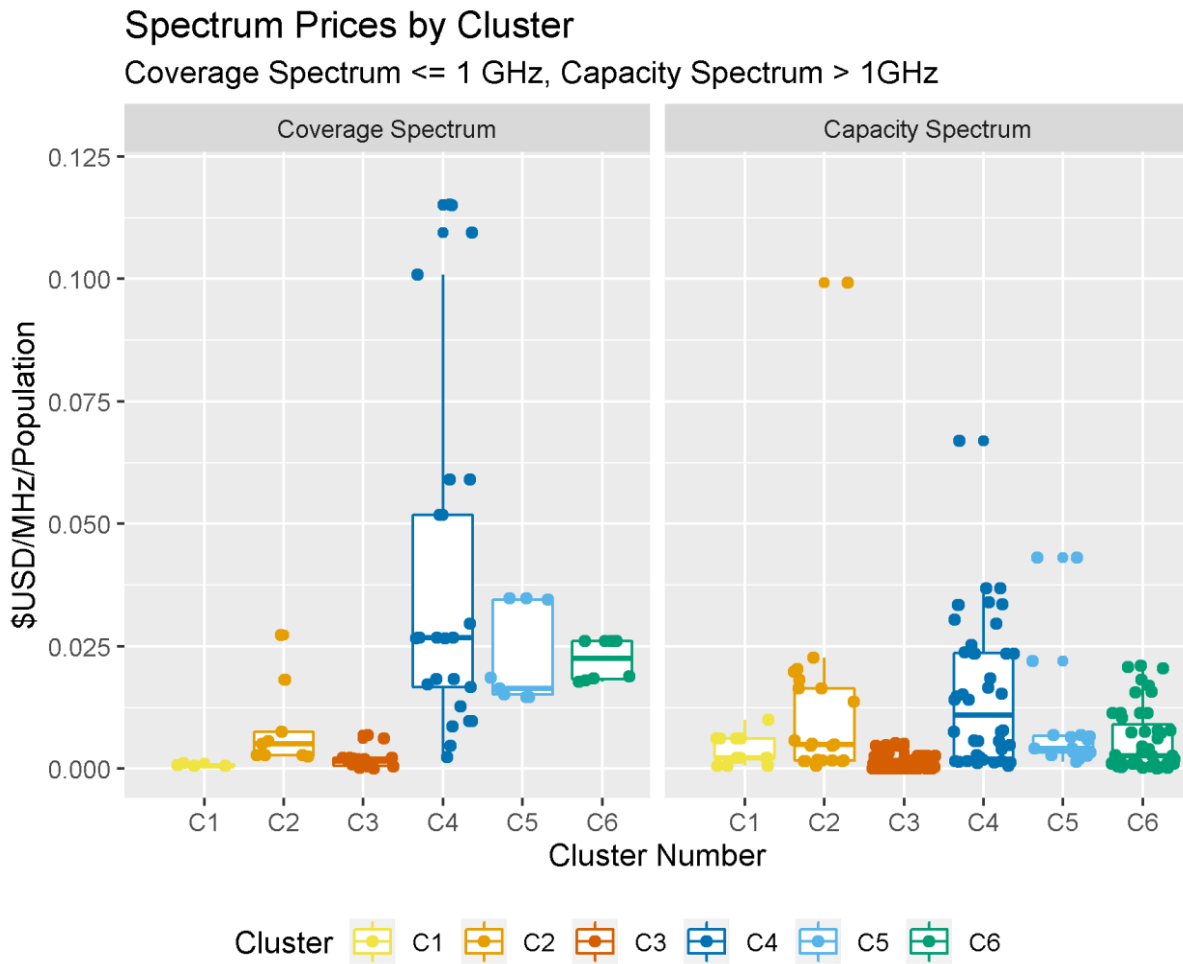

Spectrum prices are lower in the clusters with lower GDP per capita, as spectrum prices are correlated with the ARPU and relative market power. In most clusters, coverage spectrum is valued higher, except in the lowest income clusters such as C1 and C2. Generally, high spectrum prices are most closely associated with higher income countries, such as Cluster 4, 5 and 6. However, these costs are not comprehensive and do not include ongoing fees. This information is supplemented by analysis for the GSMA by NERA who analyze spectrum prices for countries around the globe (NERA Economic Consulting, 2017a), and for Latin America and the Caribbean (LAC) (NERA Economic Consulting, 2017b). As the auction prices in Figure S6 are only a fraction of the total costs, Table 3 in the paper additionally includes the Net Present Value of ongoing fees over the license duration.

To determine the prices used we have partial data for Senegal, Albania, Peru and Mexico. Senegal historically exceeds the median price of $0.13 per MHz/pop for capacity bands, suggesting very high prices relative to the market power. Hence, baseline values are selected with are close to the median

values for other countries with similar stage of development. Albania is included in NERA's analysis for capacity spectrum, but there is little information for coverage price. We treat Albania as having similar to the median price for coverage ($0.4/MHz), with the median price for the capacity ($0.1/MHz) (NERA Economic Consulting, 2017a). Peru has mixed spectrum prices, with some bands (e.g. coverage spectrum) being very high, ranging from $0.25 to $0.7 MHz/pop, compared to the median of $0.3 MHz/pop (NERA Economic Consulting, 2017b). For the baseline, both Peru and Mexico are treated as being close to the median LAC price for coverage ($0.2/MHz) and capacity ($0.1/MHz) bands. We lack data for Malawi, Uganda, Kenya and Pakistan, and therefore use similar values to other countries with comparable market power and an analogous stage of economic development identified via the clustering process to determine the prices used.

2.8. Cost calculations

Network costs ($Network_i$) are estimated based on the sum of the Radio Access Network ($RAN_i$), backhaul ($Backhaul_i$) and core ($Core_i$) components in the $i$th, as stated in equation (14):

$$Network_i = RAN_i + Backhaul_i + Core_i \quad (14)$$

Spectrum costs ($Spectrum_i$) are calculated for the $i$th area for all $f$ frequencies, given the dollar cost per MHz per capita ($Cost\_\$\_MHz\_pop_f$) multiplied by the frequency bandwidth ($Bandwidth_f$) and population ($Population_i$), as stated below in equation (15):

$$Spectrum_i = \sum_f Cost\_\$\_MHz\_pop_f * Bandwidth_f * Population_i \quad (15)$$

The quantity of tax ($Tax_i$) due for payment in the $i$th area is estimated based on the sum of network cost ($Network_i$) and spectrum cost ($Spectrum_i$) for a given tax rate ($Tax\_Rate$) which is set at 25% in the baseline, as stated in equation (16):

$$Tax_i = Network_i * \left(\frac{Tax\_Rate}{100}\right) \quad (16)$$

Finally, as digital infrastructure is provided by private network operators a profit margin ($Profit_i$) is added of 20% on all investment outgoings in the $i$th area (in addition to the WACC risk premium), as stated in equation (17):

$$Profit_i = (Network_i + Spectrum_i + Tax_i) * \left(\frac{Profit\_Margin}{100}\right) \quad (17)$$

Importantly, the modeling approach allows the MNO to take the profit margin in viable areas, but revenue beyond this quantity is reallocated to less viable areas via a user cross-subsidy.

3. Results

Figure S7 Distribution of smartphones and sites by country

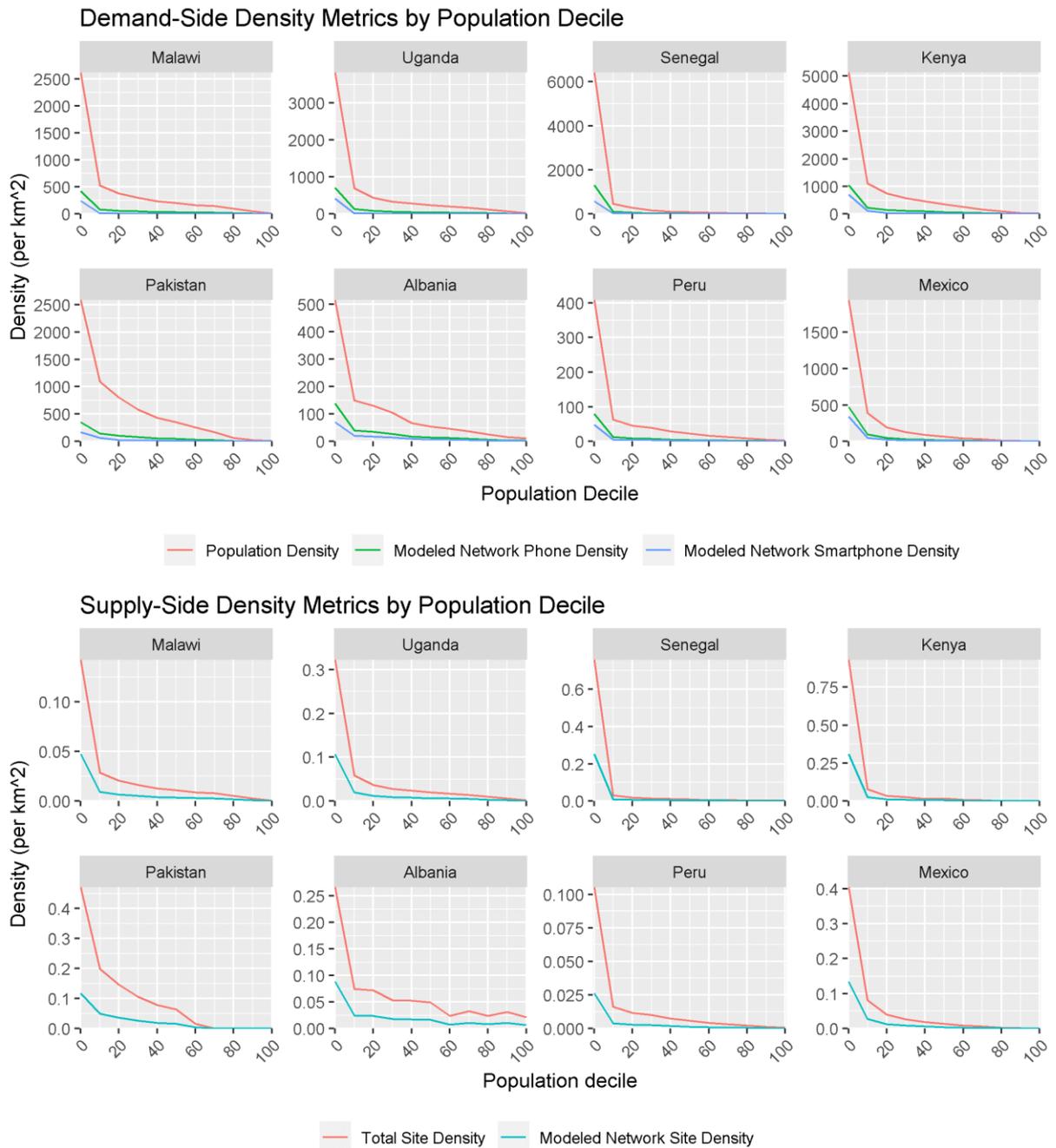

# Table S6 Cost results for all technologies, scenarios and strategies

Full Technology Results by Country

| Scenario | Strategy | Metric | C1 Malawi | Uganda | C2 Senegal | Kenya | C3 Pakistan | C4 Albania | C5 Peru | C6 Mexico |
|---|---|---|---|---|---|---|---|---|---|---|
| S1 (<25 Mbps) | 4G (W) | Government Cost ($Bn) | -0.06 | -0.21 | -0.32 | -1.1 | -2.22 | -0.09 | -2.57 | -8.97 |
| S1 (<25 Mbps) | 4G (W) | Private Cost ($Bn) | 1.12 | 1.57 | 1.26 | 4.38 | 19.89 | 0.11 | 5.42 | 14.43 |
| S1 (<25 Mbps) | 4G (W) | Social Cost ($Bn) | 1.06 | 1.36 | 0.94 | 3.28 | 17.66 | 0.02 | 2.85 | 5.46 |
| S1 (<25 Mbps) | 4G (F) | Government Cost ($Bn) | 0.29 | -0.07 | -0.38 | -1.31 | 3.88 | -0.09 | -2.78 | -9.33 |
| S1 (<25 Mbps) | 4G (F) | Private Cost ($Bn) | 1.5 | 2.88 | 1.91 | 6.67 | 26.61 | 0.12 | 7.65 | 18.3 |
| S1 (<25 Mbps) | 4G (F) | Social Cost ($Bn) | 1.78 | 2.81 | 1.53 | 5.36 | 30.49 | 0.03 | 4.87 | 8.97 |
| S1 (<25 Mbps) | 5G NSA (W) | Government Cost ($Bn) | 0.04 | -0.25 | -0.43 | -1.45 | -2.57 | -0.13 | -2.19 | -7.22 |
| S1 (<25 Mbps) | 5G NSA (W) | Private Cost ($Bn) | 1.24 | 1.63 | 1.27 | 4.33 | 14.21 | 0.31 | 8.3 | 40.93 |
| S1 (<25 Mbps) | 5G NSA (W) | Social Cost ($Bn) | 1.28 | 1.37 | 0.85 | 2.88 | 11.63 | 0.18 | 6.11 | 33.71 |
| S1 (<25 Mbps) | 5G SA (F) | Government Cost ($Bn) | 1.3 | 2.44 | -0.55 | -1.31 | 15.21 | -0.16 | -3.09 | -10.77 |
| S1 (<25 Mbps) | 5G SA (F) | Private Cost ($Bn) | 2.63 | 5.68 | 2.57 | 9.6 | 39.33 | 0.68 | 17.99 | 79.31 |
| S1 (<25 Mbps) | 5G SA (F) | Social Cost ($Bn) | 3.93 | 8.12 | 2.02 | 8.29 | 54.54 | 0.52 | 14.9 | 68.54 |
| S2 (<200 Mbps) | 4G (W) | Government Cost ($Bn) | 0.78 | 0.2 | -0.4 | -1.39 | 27.7 | -0.09 | -2.85 | -9.93 |
| S2 (<200 Mbps) | 4G (W) | Private Cost ($Bn) | 2.04 | 3.17 | 2.08 | 7.61 | 52.86 | 0.13 | 8.48 | 24.87 |
| S2 (<200 Mbps) | 4G (W) | Social Cost ($Bn) | 2.82 | 3.37 | 1.68 | 6.21 | 80.56 | 0.05 | 5.63 | 14.94 |
| S2 (<200 Mbps) | 4G (F) | Government Cost ($Bn) | 1.19 | 1.66 | -0.48 | -0.13 | 44.8 | -0.09 | -3.14 | -10.47 |
| S2 (<200 Mbps) | 4G (F) | Private Cost ($Bn) | 2.49 | 4.79 | 2.98 | 10.51 | 71.7 | 0.15 | 11.53 | 30.64 |
| S2 (<200 Mbps) | 4G (F) | Social Cost ($Bn) | 3.68 | 6.45 | 2.5 | 10.37 | 116.5 | 0.06 | 8.39 | 20.18 |
| S2 (<200 Mbps) | 5G NSA (W) | Government Cost ($Bn) | 0.37 | -0.09 | -0.5 | -1.65 | 14.43 | -0.13 | -2.44 | -8.62 |
| S2 (<200 Mbps) | 5G NSA (W) | Private Cost ($Bn) | 1.6 | 2.9 | 2.1 | 6.51 | 38.47 | 0.37 | 10.95 | 56.07 |
| S2 (<200 Mbps) | 5G NSA (W) | Social Cost ($Bn) | 1.97 | 2.81 | 1.6 | 4.85 | 52.9 | 0.24 | 8.51 | 47.45 |
| S2 (<200 Mbps) | 5G SA (F) | Government Cost ($Bn) | 1.73 | 4.69 | 0.22 | 1.21 | 137.02 | -0.17 | 4.52 | 12.9 |
| S2 (<200 Mbps) | 5G SA (F) | Private Cost ($Bn) | 3.1 | 8.16 | 3.94 | 12.38 | 173.55 | 0.75 | 28.43 | 109.51 |
| S2 (<200 Mbps) | 5G SA (F) | Social Cost ($Bn) | 4.83 | 12.85 | 4.16 | 13.58 | 310.57 | 0.58 | 32.95 | 122.41 |
| S3 (<400 Mbps) | 4G (W) | Government Cost ($Bn) | 1.84 | 1.99 | -0.42 | 1.22 | 66.12 | -0.11 | -3.23 | -10.34 |
| S3 (<400 Mbps) | 4G (W) | Private Cost ($Bn) | 3.21 | 5.15 | 3.12 | 12 | 95.2 | 0.36 | 12.55 | 29.31 |
| S3 (<400 Mbps) | 4G (W) | Social Cost ($Bn) | 5.04 | 7.14 | 2.7 | 13.22 | 161.32 | 0.25 | 9.32 | 18.96 |
| S3 (<400 Mbps) | 4G (F) | Government Cost ($Bn) | 2.29 | 3.91 | 0.59 | 4.29 | 87.65 | -0.12 | -3.65 | -11.01 |
| S3 (<400 Mbps) | 4G (F) | Private Cost ($Bn) | 3.71 | 7.27 | 4.23 | 15.38 | 118.91 | 0.44 | 17.11 | 36.48 |
| S3 (<400 Mbps) | 4G (F) | Social Cost ($Bn) | 6 | 11.18 | 4.82 | 19.67 | 206.56 | 0.32 | 13.46 | 25.47 |
| S3 (<400 Mbps) | 5G NSA (W) | Government Cost ($Bn) | 0.67 | 0.47 | -0.53 | -1.78 | 27.96 | -0.14 | -2.66 | -9.16 |
| S3 (<400 Mbps) | 5G NSA (W) | Private Cost ($Bn) | 1.94 | 3.51 | 2.37 | 7.86 | 53.38 | 0.45 | 13.41 | 61.88 |
| S3 (<400 Mbps) | 5G NSA (W) | Social Cost ($Bn) | 2.61 | 3.97 | 1.84 | 6.08 | 81.34 | 0.31 | 10.75 | 52.72 |
| S3 (<400 Mbps) | 5G SA (F) | Government Cost ($Bn) | 2.14 | 5.5 | 0.57 | 2.8 | 253.11 | -0.18 | 20.17 | 23.24 |
| S3 (<400 Mbps) | 5G SA (F) | Private Cost ($Bn) | 3.56 | 9.05 | 4.32 | 14.13 | 301.47 | 0.86 | 45.68 | 120.91 |
| S3 (<400 Mbps) | 5G SA (F) | Social Cost ($Bn) | 5.7 | 14.55 | 4.89 | 16.92 | 554.59 | 0.68 | 65.85 | 144.14 |

[1] Infrastructure Sharing Strategy: Baseline.
[2] Spectrum Pricing Strategy: Baseline.
[3] Taxation Strategy: Baseline.

Table S7 Cost results for infrastructure sharing strategies and scenarios

Infrastructure Sharing Results by Country

| Scenario | Strategy | Metric | C1 Malawi | Uganda | C2 Senegal | Kenya | C3 Pakistan | C4 Albania | C5 Peru | C6 Mexico |
|---|---|---|---|---|---|---|---|---|---|---|
| S1 (<25 Mbps) | Baseline | Social Cost ($Bn) | 1.28 | 1.37 | 0.85 | 2.88 | 11.63 | 0.18 | 6.11 | 33.71 |
| S1 (<25 Mbps) | Baseline | Private Cost ($Bn) | 1.24 | 1.63 | 1.27 | 4.33 | 14.21 | 0.31 | 8.3 | 40.93 |
| S1 (<25 Mbps) | Baseline | Government Cost ($Bn) | 0.04 | -0.25 | -0.43 | -1.45 | -2.57 | -0.13 | -2.19 | -7.22 |
| S1 (<25 Mbps) | Passive | Social Cost ($Bn) | 1.08 | 1.2 | 0.71 | 2.41 | 9.06 | 0.17 | 5.04 | 28.66 |
| S1 (<25 Mbps) | Passive | Private Cost ($Bn) | 1.14 | 1.44 | 1.12 | 3.81 | 11.37 | 0.29 | 7.12 | 35.37 |
| S1 (<25 Mbps) | Passive | Government Cost ($Bn) | -0.05 | -0.24 | -0.41 | -1.4 | -2.31 | -0.13 | -2.08 | -6.71 |
| S1 (<25 Mbps) | Active | Social Cost ($Bn) | 0.71 | 0.73 | 0.37 | 1.16 | 3.36 | 0.1 | 1.94 | 15 |
| S1 (<25 Mbps) | Active | Private Cost ($Bn) | 0.83 | 0.91 | 0.75 | 2.44 | 5.09 | 0.22 | 3.7 | 20.32 |
| S1 (<25 Mbps) | Active | Government Cost ($Bn) | -0.12 | -0.19 | -0.38 | -1.28 | -1.73 | -0.12 | -1.77 | -5.31 |
| S1 (<25 Mbps) | SRN | Social Cost ($Bn) | 0.36 | 0.46 | 0.28 | 0.96 | 2.91 | 0.06 | 1.53 | 11.23 |
| S1 (<25 Mbps) | SRN | Private Cost ($Bn) | 0.45 | 0.62 | 0.65 | 2.22 | 4.59 | 0.18 | 3.25 | 16.16 |
| S1 (<25 Mbps) | SRN | Government Cost ($Bn) | -0.09 | -0.16 | -0.37 | -1.26 | -1.68 | -0.11 | -1.72 | -4.93 |
| S2 (<200 Mbps) | Baseline | Social Cost ($Bn) | 1.97 | 2.81 | 1.6 | 4.85 | 52.9 | 0.24 | 8.51 | 47.45 |
| S2 (<200 Mbps) | Baseline | Private Cost ($Bn) | 1.6 | 2.9 | 2.1 | 6.51 | 38.47 | 0.37 | 10.95 | 56.07 |
| S2 (<200 Mbps) | Baseline | Government Cost ($Bn) | 0.37 | -0.09 | -0.5 | -1.65 | 14.43 | -0.13 | -2.44 | -8.62 |
| S2 (<200 Mbps) | Passive | Social Cost ($Bn) | 1.62 | 2.04 | 1.24 | 3.87 | 33.63 | 0.21 | 6.77 | 38.3 |
| S2 (<200 Mbps) | Passive | Private Cost ($Bn) | 1.42 | 2.36 | 1.71 | 5.42 | 28.37 | 0.34 | 9.03 | 45.98 |
| S2 (<200 Mbps) | Passive | Government Cost ($Bn) | 0.2 | -0.32 | -0.47 | -1.55 | 5.26 | -0.13 | -2.26 | -7.69 |
| S2 (<200 Mbps) | Active | Social Cost ($Bn) | 0.81 | 1.11 | 0.62 | 1.82 | 8.87 | 0.12 | 2.54 | 19.58 |
| S2 (<200 Mbps) | Active | Private Cost ($Bn) | 0.95 | 1.34 | 1.02 | 3.17 | 11.16 | 0.24 | 4.37 | 25.36 |
| S2 (<200 Mbps) | Active | Government Cost ($Bn) | -0.13 | -0.23 | -0.4 | -1.34 | -2.29 | -0.12 | -1.83 | -5.78 |
| S2 (<200 Mbps) | SRN | Social Cost ($Bn) | 0.47 | 0.84 | 0.53 | 1.62 | 8.42 | 0.08 | 2.13 | 15.81 |
| S2 (<200 Mbps) | SRN | Private Cost ($Bn) | 0.57 | 1.04 | 0.93 | 2.94 | 10.66 | 0.2 | 3.91 | 21.21 |
| S2 (<200 Mbps) | SRN | Government Cost ($Bn) | -0.1 | -0.2 | -0.4 | -1.32 | -2.24 | -0.12 | -1.79 | -5.4 |
| S3 (<400 Mbps) | Baseline | Social Cost ($Bn) | 2.61 | 3.97 | 1.84 | 6.08 | 81.34 | 0.31 | 10.75 | 52.72 |
| S3 (<400 Mbps) | Baseline | Private Cost ($Bn) | 1.94 | 3.51 | 2.37 | 7.86 | 53.38 | 0.45 | 13.41 | 61.88 |
| S3 (<400 Mbps) | Baseline | Government Cost ($Bn) | 0.67 | 0.47 | -0.53 | -1.78 | 27.96 | -0.14 | -2.66 | -9.16 |
| S3 (<400 Mbps) | Passive | Social Cost ($Bn) | 2.11 | 2.73 | 1.44 | 4.84 | 54.07 | 0.28 | 8.43 | 42.35 |
| S3 (<400 Mbps) | Passive | Private Cost ($Bn) | 1.68 | 2.85 | 1.92 | 6.49 | 39.08 | 0.42 | 10.85 | 50.45 |
| S3 (<400 Mbps) | Passive | Government Cost ($Bn) | 0.43 | -0.13 | -0.49 | -1.65 | 14.99 | -0.14 | -2.43 | -8.1 |
| S3 (<400 Mbps) | Active | Social Cost ($Bn) | 0.9 | 1.29 | 0.7 | 2.24 | 12.25 | 0.15 | 3.1 | 21.31 |
| S3 (<400 Mbps) | Active | Private Cost ($Bn) | 1.04 | 1.53 | 1.11 | 3.62 | 14.89 | 0.27 | 4.98 | 27.27 |
| S3 (<400 Mbps) | Active | Government Cost ($Bn) | -0.14 | -0.25 | -0.41 | -1.39 | -2.64 | -0.12 | -1.88 | -5.96 |
| S3 (<400 Mbps) | SRN | Social Cost ($Bn) | 0.57 | 1.03 | 0.61 | 2.03 | 11.8 | 0.1 | 2.69 | 17.57 |
| S3 (<400 Mbps) | SRN | Private Cost ($Bn) | 0.68 | 1.25 | 1.02 | 3.39 | 14.39 | 0.22 | 4.53 | 23.15 |
| S3 (<400 Mbps) | SRN | Government Cost ($Bn) | -0.11 | -0.22 | -0.4 | -1.37 | -2.59 | -0.12 | -1.84 | -5.58 |

[1] Technology Strategy: 5G NSA with Wireless Backhaul.
[2] Spectrum Pricing Strategy: Baseline.
[3] Taxation Strategy: Baseline.

# Table S8 Cost results for spectrum pricing strategies and scenarios

Spectrum Pricing Results by Country

| Scenario | Strategy | Metric | C1 Malawi | Uganda | C2 Senegal | Kenya | C3 Pakistan | C4 Albania | C5 Peru | C6 Mexico |
|---|---|---|---|---|---|---|---|---|---|---|
| S1 (<25 Mbps) | Low Prices (-75%) | Social Cost ($Bn) | 1.24 | 1.37 | 0.85 | 2.88 | 11.63 | 0.18 | 6.11 | 33.71 |
| S1 (<25 Mbps) | Low Prices (-75%) | Private Cost ($Bn) | 1.2 | 1.54 | 1.02 | 3.47 | 13.17 | 0.23 | 7.12 | 38.09 |
| S1 (<25 Mbps) | Low Prices (-75%) | Government Cost ($Bn) | 0.04 | -0.17 | -0.17 | -0.58 | -1.53 | -0.05 | -1.01 | -4.38 |
| S1 (<25 Mbps) | Baseline | Social Cost ($Bn) | 1.28 | 1.37 | 0.85 | 2.88 | 11.63 | 0.18 | 6.11 | 33.71 |
| S1 (<25 Mbps) | Baseline | Private Cost ($Bn) | 1.24 | 1.63 | 1.27 | 4.33 | 14.21 | 0.31 | 8.3 | 40.93 |
| S1 (<25 Mbps) | Baseline | Government Cost ($Bn) | 0.04 | -0.25 | -0.43 | -1.45 | -2.57 | -0.13 | -2.19 | -7.22 |
| S1 (<25 Mbps) | High Prices (+100%) | Social Cost ($Bn) | 1.33 | 1.37 | 0.85 | 2.88 | 11.63 | 0.18 | 6.11 | 33.71 |
| S1 (<25 Mbps) | High Prices (+100%) | Private Cost ($Bn) | 1.29 | 1.74 | 1.62 | 5.49 | 15.59 | 0.42 | 9.86 | 44.71 |
| S1 (<25 Mbps) | High Prices (+100%) | Government Cost ($Bn) | 0.04 | -0.37 | -0.77 | -2.61 | -3.96 | -0.24 | -3.76 | -11 |
| S2 (<200 Mbps) | Low Prices (-75%) | Social Cost ($Bn) | 1.93 | 2.72 | 1.6 | 4.85 | 51.86 | 0.24 | 8.51 | 47.45 |
| S2 (<200 Mbps) | Low Prices (-75%) | Private Cost ($Bn) | 1.57 | 2.81 | 1.85 | 5.64 | 37.43 | 0.29 | 9.77 | 53.23 |
| S2 (<200 Mbps) | Low Prices (-75%) | Government Cost ($Bn) | 0.37 | -0.09 | -0.25 | -0.78 | 14.43 | -0.05 | -1.26 | -5.78 |
| S2 (<200 Mbps) | Baseline | Social Cost ($Bn) | 1.97 | 2.81 | 1.6 | 4.85 | 52.9 | 0.24 | 8.51 | 47.45 |
| S2 (<200 Mbps) | Baseline | Private Cost ($Bn) | 1.6 | 2.9 | 2.1 | 6.51 | 38.47 | 0.37 | 10.95 | 56.07 |
| S2 (<200 Mbps) | Baseline | Government Cost ($Bn) | 0.37 | -0.09 | -0.5 | -1.65 | 14.43 | -0.13 | -2.44 | -8.62 |
| S2 (<200 Mbps) | High Prices (+100%) | Social Cost ($Bn) | 2.02 | 2.92 | 1.6 | 4.85 | 54.29 | 0.24 | 8.51 | 47.45 |
| S2 (<200 Mbps) | High Prices (+100%) | Private Cost ($Bn) | 1.65 | 3.01 | 2.44 | 7.67 | 39.86 | 0.48 | 12.52 | 59.85 |
| S2 (<200 Mbps) | High Prices (+100%) | Government Cost ($Bn) | 0.37 | -0.09 | -0.85 | -2.81 | 14.43 | -0.24 | -4 | -12.4 |
| S3 (<400 Mbps) | Low Prices (-75%) | Social Cost ($Bn) | 2.58 | 3.89 | 1.84 | 6.08 | 80.3 | 0.31 | 10.75 | 52.72 |
| S3 (<400 Mbps) | Low Prices (-75%) | Private Cost ($Bn) | 1.9 | 3.42 | 2.11 | 6.99 | 52.34 | 0.37 | 12.23 | 59.04 |
| S3 (<400 Mbps) | Low Prices (-75%) | Government Cost ($Bn) | 0.67 | 0.47 | -0.27 | -0.91 | 27.96 | -0.06 | -1.49 | -6.32 |
| S3 (<400 Mbps) | Baseline | Social Cost ($Bn) | 2.61 | 3.97 | 1.84 | 6.08 | 81.34 | 0.31 | 10.75 | 52.72 |
| S3 (<400 Mbps) | Baseline | Private Cost ($Bn) | 1.94 | 3.51 | 2.37 | 7.86 | 53.38 | 0.45 | 13.41 | 61.88 |
| S3 (<400 Mbps) | Baseline | Government Cost ($Bn) | 0.67 | 0.47 | -0.53 | -1.78 | 27.96 | -0.14 | -2.66 | -9.16 |
| S3 (<400 Mbps) | High Prices (+100%) | Social Cost ($Bn) | 2.66 | 4.09 | 1.84 | 6.13 | 82.73 | 0.31 | 10.75 | 52.72 |
| S3 (<400 Mbps) | High Prices (+100%) | Private Cost ($Bn) | 1.99 | 3.62 | 2.71 | 9.02 | 54.77 | 0.56 | 14.98 | 65.66 |
| S3 (<400 Mbps) | High Prices (+100%) | Government Cost ($Bn) | 0.67 | 0.47 | -0.87 | -2.89 | 27.96 | -0.25 | -4.23 | -12.94 |

[1] Technology Strategy: 5G NSA with Wireless Backhaul.

[2] Infrastructure Sharing Strategy: Baseline.

[3] Taxation Strategy: Baseline.